\documentclass[a4paper,aps,pra,twocolumn,amsmath,amssymb,nofootinbib,showpacs,superscriptaddress,accepted=2023-10-12]{quantumarticle}
\pdfoutput=1
\usepackage[english]{babel}
\usepackage{latexsym}
\usepackage{graphics}
\usepackage[demo]{graphicx}
\usepackage{epsfig}
\usepackage[dvipsnames]{xcolor}
\usepackage{bm}
\usepackage{amsmath}
\usepackage{amssymb}
\usepackage{amsthm}
\usepackage{dcolumn}
\usepackage{bm}
\usepackage{subfig}
\usepackage{float}
\usepackage{tikz} 
\usepackage{soul}
\usetikzlibrary{calc}
\usepackage{pgfplots}
\pgfplotsset{compat=newest, width=8 cm}
\usetikzlibrary{arrows.meta,positioning}
\usetikzlibrary{decorations.markings}
\usepgfplotslibrary{fillbetween}
\usepgfplotslibrary{polar}
\usepackage{pgfplotstable}
\usepackage{hyperref}
\graphicspath{{Images/}}
\usepackage{comment}
\usepackage{epstopdf}
\usepackage{cancel}
\usepackage{cleveref}
\usepackage{enumerate}
\usepackage{braket}
\usepackage{appendix}

\newcommand{\bs}{\boldsymbol}
\newcommand{\ovl}{\overline}

\theoremstyle{remark}

\tikzset{
    set arrow inside/.code={\pgfqkeys{/tikz/arrow inside}{#1}},
    set arrow inside={end/.initial=>, opt/.initial=},
    /pgf/decoration/Mark/.style={
        mark/.expanded=at position #1 with
        {
            \noexpand\arrow[\pgfkeysvalueof{/tikz/arrow inside/opt}]{\pgfkeysvalueof{/tikz/arrow inside/end}}
        }
    },
    arrow inside/.style 2 args={
        set arrow inside={#1},
        postaction={
            decorate,decoration={
                markings,Mark/.list={#2}
            }
        }
    },
}

\begin{document}

\title{Mean field study of 2D quasiparticle condensate formation in presence of strong decay}
\author{N.A. Asriyan}
\email{naasriyan@vniia.ru}
\affiliation{N.L. Dukhov Research Institute of Automatics (VNIIA), Moscow 127030, Russia;}

\author{A.A. Elistratov}
\affiliation{N.L. Dukhov Research Institute of Automatics (VNIIA), Moscow 127030, Russia;}

\author{Yu.E. Lozovik}
\affiliation{Institute for Spectroscopy RAS, Troitsk 108840, Moscow, Russia;}
\affiliation{Moscow Institute of Electronics and Mathematics, National Research University Higher School of Economics, 101000 Moscow,
Russia;}
\begin{abstract}
 Bose-condensation in a system of 2D quasiparticles is considered in the scope of a microscopic model. Mean-field dynamical equations are derived with the help of the Schwinger-Keldysh formalism and a simple model is proposed which allows to describe key features of condensate formation in systems with various quasiparticle decay rates. By analysing stationary solutions of this equation, we obtain the phase diagram of quasiparticle gas, finding a bistability region in the parameter space of the system. Finally, as an application of our theory, we study the phase diagram of a 2D exciton-polariton system in CdTe microcavity.
\end{abstract}

\maketitle
\section{Introduction}

Decades after Bose-Einstein condensation was predicted theoretically~\cite{Bose1924, Einstein1925}, it was observed directly in experiments with cold atoms in 1995~\cite{Anderson, Bradley, Davis}. This observation was followed by discovering bose-condensates in many other systems such as quantum well excitons~\cite{HighNature2012, HighNanolett2012},  exciton-polaritons~\cite{Kasprzak, Balili},  magnons~\cite{Demokritov} and microcavity photons~\cite{Klaers}.


Though during condensate formation the cold atom gas is out of equilibrium, the resulting condensate state is an equilibrium one, which is not the case for solid state quasiparticle systems. Despite the similar nature of the low-temperature state in these systems and in cold atomic gases, the quasiparticle condensate is different in several aspects. Firstly, due to finite lifetime, these systems need to be pumped externally, hence the condensate is in a quasi-equilibrium state which is determined by an interplay between pumping and decay processes. Moreover, these condensates are often considered in low dimensional systems and together with small masses of quasiparticles, it may change totally the relevant scales (energies, times) of the condensate and the process of its formation. One of the most attractive features is the potential ability to observe high temperature condensation.

The basic Gross-Pitaevskii equation, describing condensate in equilibrium systems has been modified in numerous ways (leading to dissipative Gross–Pitaevskii-type models) in order to describe phenomenologically non-equilibrium physics of exciton/photon/exciton-polariton condensates. To step beyond the description of the kinetic stage of condensate formation, which had been well studied ~\cite{Imamogluprla,Piermarocchi1996, Stenius} and to incorporate the coherent properties of condensate in the evolution equation, condensate is commonly considered to be an open quantum system subject to reservoirs, namely the pump reservoir, decay reservoir, lattice phonon reservoir, etc.

Several phenomenological models were proposed in this scope. Namely, the most common is the one introduced by M. Wouters and I. Carusotto\cite{Wouters2007}, which describes polariton condensate as a system, coupled to classical excitonic reservoir of density $n_{\rm R}$:
\begin{equation}
    \begin{cases}
        i \dot\psi=\left\{-\frac{\hbar^2\nabla^{2}}{2 m}+\frac{i}{2}\left[R\left(n_{R}\right)-\gamma\right]+g|\psi|^{2}+2 \tilde{g} n_{R}\right\} \psi,\\
        \dot n_{R}=P-\gamma_{R} n_{R}-R\left(n_{R}\right)|\psi(\mathbf{r})|^{2}+D \nabla^{2} n_{R}.
    \end{cases}\label{eq:Wouters_dynamics}
\end{equation}
Here $P$ describes incoherent pumping, $R(n_{R})$ is an amplification rate of reservoir-condensate scattering, $g$ and $\tilde g$ stand for intracondensate and condensate-reservoir particle interaction. Parameters $\gamma_{R}$ and $\gamma$ are decay rates of quasiparticles from reservoir and condensate.

Gain saturation, which is essential for describing condensate density equilibrating, may be incorporated  directly into the condensate evolution equation:
\begin{align}\label{eq:ODGPE}
    i \hbar \dot \psi{=}\left[{-}\frac{\hbar^{2} \nabla^{2}}{2 m}{+}g|\psi|^{2}{+}i\left(\gamma_{\mathrm{eff}}{-}\Gamma_{\mathrm{eff}}|\psi|^{2}\right)\right] \psi.
\end{align}

These appeared to be fruitful approaches, which allowed to describe spontaneous vortex lattice formation~\cite{KeelingBerloff08}, pattern formation~\cite{ManniLago2011}, as well as relaxation oscillations were considered~\cite{Opala2018}.

Models of the type \eqref{eq:Wouters_dynamics} and \eqref{eq:ODGPE} are well-suited for describing long-lifetime systems with moderate decay rates $\gamma\ll\gamma_R$. Though, with the help of \eqref{eq:Wouters_dynamics} the opposite case ($\gamma\gg\gamma_R$) was also considered ~\cite{Smirnov2014, Baboux2018}, where the modulational instability of the homogeneous condensate was demonstrated to be a consequence of slow reservoir relaxation.

However, the above-mentioned models share several drawbacks. They do not allow considering the normal phase-condensate transition in systems with wide range of decay rates of quasiparticles: from the "ultracold atom gas" limit with negligible decay to the "strong dissipation" case in presence of high losses from condensate compensated by high gain rates from the reservoir. For the latter case, one should consider the impact of the broadening of the condensate spectral function on the interaction with the reservoir, which introduces memory effects.

As we will show further, the key to overcoming this issue is considering the frequency-dependent gain. One way to do this is using a phenomenological non-Hermitian term first introduced by L.P. Pitaevskii~\cite{Pitaevski1959} for superfluid He II and later adopted for describing frequency dependent gain due to the polariton-polariton interaction~\cite{Wouters2010}:
\begin{align}
    i \hbar \dot\psi{=}\left[{-}\frac{\hbar^{2} \nabla^{2}}{2 m}{+}g|\psi|^{2}{+}iP\left(1{-}\frac{i\partial_t}{\Omega_{\rm tr}}
    \right){+}i\left(\gamma{-}\Gamma|\psi|^{2}\right)\right] \psi.
\end{align}
Here, close to the threshold frequency $\Omega_{\rm tr}$ the gain efficiency decreases to zero. Using such a model for polariton condensate appeared to be essential for obtaining its excitation spectrum and describing its superfluid properties, as shown in ~\cite{Wouters2010}.

Along with studying the model equations, microscopic theories were developed for exciton/exciton polariton systems~\cite{Haug2014, Elistratov2018} in order to derive evolution equations \textit{ab initio}. This activity resulted in deriving microscopically motivated expressions for the terms of equations \eqref{eq:Wouters_dynamics} and \eqref{eq:ODGPE}. {However, a similar microscopical treatment, which could reproduce the frequency-dependent gain and could be applicable in presence of strong decay, is lacking}.

In this paper, we propose a microscopically motivated equation for condensate formation dynamics in 2D finite-lifetime quasiparticle system, which allows considering consistently the cases of various decay rates due to a properly described frequency-dependent gain effect. Stated otherwise, the approach we use aims to be applicable to both the case of long-living particles (i.e. weakly coupled to the decay bath) and rapidly-decaying quasiparticles whose lifetime may be much smaller than reservoir evolution time scales. In the latter case the reservoir acts as a system with long memory effect providing a delayed feedback on condensate evolution, i.e. it may be considered as a non-Markovian evolution regime. Using this dynamical equation, we obtain the mean-field phase diagram for a quasiparticle system and demonstrate that finite lifetime of condensate particles may lead to formation of condensate and normal phase overlap regions on the phase diagram. Moreover, we show that above-mentioned memory effects significantly affect the condensate dynamics leading to evolution patterns different from exponential approach to equilibrium and density relaxation oscillations. As a demonstration of real-life application of the dynamical equation we derive, we will consider the phase diagram for CdTe microcavity polariton gas.

Being motivated by the discussion of spontaneous symmetry breaking in cold atom condensate by H. T. C. Stoof~\cite{Stoof_coher, Stoof1999}, we use the same theoretical framework: the Schwinger-Keldysh technique in path integral formulation. It provides direct access to the condensate order parameter, that's why it has been widely applied to quasiparticle condensates. For instance, photonic condensate in a dye-filled optical microcavity ~\cite{Leeuw2013}, exciton polariton condensates in quantum wells~\cite{Haug2014, Elistratov2018}, parametrically pumped polariton systems ~\cite{Dunnett2016} were also considered. We use here a single-level condensate model (e.g. assume uniform condensate in a finite-sized system) in the spirit of pioneering articles~\cite{Imamogluprla, Stenius} dealing with exciton condensation. Of course, this prevents us from studying the spatial structure of the condensed state, namely from reproducing the results of ~\cite{Baboux2018}. However, this simplification allows keeping the further discussion analytically tractable and focusing on the impact of intense condensate decay on the evolution equations.


We start from defining in Sec. \ref{sec:model} the model Hamiltonian for 2D quasiparticles with pair interaction. Here our goal is to develop a general theory, not considering any particular system. Introducing the necessary parameters, we derive the mean-field dynamical equations for condensate evolution in Sec. \ref{sec:model_equations}. This derivation is followed by a description of a possible simplified approximate expressions for equation terms. It leads to a dynamical model, which is one of the main results of the current paper.

The next Sec. \ref{sec:phdiag} is devoted to discussion of a mean-field phase diagram for a generic quasiparticle system. Possible phases are described followed by a stability analysis. A detailed description of condensation dynamics is presented in Sec. \ref{sec:dyneffects}.

After a discussion on the advantages and drawbacks of the model in Sec. \ref{sec:discussion}, in the last Sec. \ref{sec:demonstration} of the article we present an application of the model to real-life system such as exciton polariton gas in CdTe/CdMgTe microcavity.

\section{The model system}\label{sec:model}
We deal with a 2D single-level condensate model of finite-lifetime quasiparticles embedded in a long-living particle reservoir. The system is treated as a 2D bose gas with contact interparticle interaction and leakage from the condensate. The model Hamiltonian for the system described is as follows ($\hat\psi_{\bs q}$ is the quasiparticle annihilation operator with wavevector $\bs q$, therefore $\hat\psi_0$ corresponds to the condensate mode):
\begin{align}\label{eq:Hamiltonian}
    \hat H = \left(\varepsilon-i\frac{\Gamma}{2}\right)\hat \psi_0^{\dagger}\hat \psi_0+&\sum_{\bs q\neq\bs 0}\varepsilon_{\bs q}\hat \psi_{\bs q}^{\dagger}\hat \psi_{\bs q}\nonumber\\
    +g_0 &\sum_{\boldsymbol{q}_{1}, \boldsymbol{q}_{2}, \boldsymbol{q}^{\prime}} \hat{\psi}_{\boldsymbol{q}_{1}{+}\bs q^{\prime}}^{\dagger} \hat{\psi}_{\boldsymbol{q}_{2}{-}\boldsymbol{q}^{\prime}}^{\dagger} \hat{\psi}_{\boldsymbol{q}_{1}} \hat{\psi}_{\boldsymbol{q}_{2}}
\end{align}
Here $g_0=\frac{V_0}{2L^2}$ stands for contact interparticle interaction with $V_0$ being the interaction potential and $L^2$ denoting the quantization area. $\varepsilon$ denotes the possible energy detuning of the condensate level with respect to the reservoir dispersion curve $\varepsilon_{\bs q\to 0}=0$. A decay rate $\Gamma$ is introduced to describe condensate particle finite lifetime. Hereafter $\hbar=1$.

The action for this system defined on the Schwinger-Keldysh contour in path integral formulation is as follows:
\begin{align}
    S&{=}\int\limits_c d\tau\left[\ovl{\psi}_0\left\{i\partial_{\tau}{-}\varepsilon{+}i\frac{\Gamma}{2}\right\}\psi_0{-}\sum_{\bs q\neq \bs 0}\ovl{\psi}_{\bs q}\left\{i\partial_{\tau}{-}\varepsilon_{\bs q}\right\}\psi_{\bs q}\right]\nonumber\\
    &{-}g_0\int\limits_c d\tau\sum_{\bs q_1, \bs q_2, \bs q'}\ovl \psi_{\bs q_1+\bs q'}\ovl\psi_{\bs q_2-\bs q'}\psi_{\bs q_1}\psi_{\bs q_2}.
\end{align}

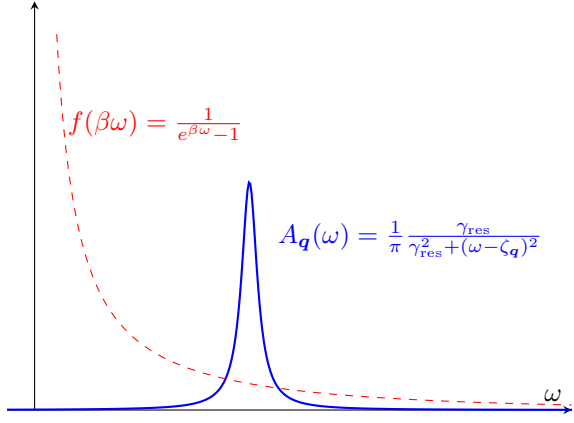
\begin{figure}[H]
	\centering
	\begin{tikzpicture}[scale=1]
            \begin{axis}[
            	axis x line = center,
            	axis y line = center,
	            xmin=-0.2, xmax=4, ymin=0, ymax=2.15,
	            xtick=\empty,
	            ytick=\empty,
	            height=7cm, width=9cm, grid=none,
	            xlabel={$\omega$},
	        ]
	        \addplot[samples=1000, blue, thick]{1.2/(200*(x-1.6)^2+1)};
	        \addplot[samples=1000, red, dashed, domain = 0.165:4]{0.17/(e^(0.5*x)-1)};
	        \node[red] at (axis cs:0.9, 1.5) {$f(\beta\omega)=\frac1{e^{\beta\omega}-1}$}; 
	        \node[blue] at (axis cs: 2.8, 0.9) {$A_{\bs q}(\omega) = \frac{1}{\pi}\frac{\gamma_{\rm res}}{\gamma^2_{\rm res}{+}(\omega-\zeta_{\bs q})^2}$};
	        \end{axis}
	\end{tikzpicture}
	\caption{The necessary assumption of “narrow” spectral function. The red and blue lines schematically depict the frequency dependencies of the bosonic distribution function and the spectral function of a reservoir quasiparticle, respectively.}
	\label{fig:HF}
\end{figure}

Our goal is to integrate out the reservoir degrees of freedom to derive an effective action for the condensate. We are going to treat the reservoir in the simplest possible way as a continuously pumped quasi-equilibrium system with stationary surface density $n$ and effective temperature $T=(k_{B}\beta)^{-1}$. Moreover, we assume the collision broadening for the reservoir to be negligible compared to the effective temperature. When considering  the greater/lesser components of the Green's function on the Schwinger-Keldysh contour, this allows to use the following approximation (the spectral function is assumed to be a sharply peaked Lorentzian, see Fig. \ref{fig:HF}):
\begin{align}
    &\frac{i G_{\boldsymbol{q}}^{>}(\omega)}{2\pi}{=}A_{\boldsymbol{q}}(\omega)\left[1{+}f(\beta\omega)\right]{\to} A_{\boldsymbol{q}}(\omega)\left[1{+}f(\beta(\zeta_{\bs q}))\right], \nonumber\\
&\frac{i G_{\boldsymbol{q}}^{<}(\omega)}{2\pi}{=}A_{\boldsymbol{q}}(\omega) f(\beta\omega){\to} A_{\boldsymbol{q}}(\omega)f(\beta(\zeta_{q})).
\end{align}
Here $f(x)=\left(e^{x}-1\right)^{-1}$, $\zeta_{\bs q}=\varepsilon_{\bs q}{+}\Delta \varepsilon_{\bs q}{-}\mu$ with $\Delta \varepsilon_{\bs q}$ taking into account the blueshift due to reservoir interparticle interaction.
And $\mu$ stands for the ideal gas chemical potential of the reservoir. From the expressions above, we may combine the casual Green's function:
\begin{align}\label{eq:reservoir_green_decomposition}
    iG^{\rm res}_{\bs q}&(t,t')=e^{-\gamma_{\rm res}|t-t'|-i\zeta_{\bs q} (t-t
')}\nonumber\\
&\times\left\{\Theta(t,t')(1+f(\zeta_{\bs q}))+\Theta(t',t)f(\zeta_{\bs q})\right\}
\end{align}
with $\gamma_{\rm res}\to 0$.

Note that the assumptions introduced in this section (the form of the spectral function with uniform broadening of all the states as well as $\gamma_{\rm res}\to 0$) are reasonable for the case of weak interparticle interaction in the reservoir. For real systems, this corresponds to weakly interacting gases. For instance, it is the case for exciton-polariton systems at reasonable temperatures. A rough estimate is $\frac{V_0n^{2D}}{kT}\ll1$. For typical excitonic densities of order $n^{\rm 2D}\sim 10^{10}$ cm${}^{-2}$ this leads to $T\gg 2\mu K$.
\section{Deriving dynamical equations}\label{sec:model_equations}
\subsection{Effective action}
\begin{figure}[H]
    \centering
    \includegraphics[width=5 cm]{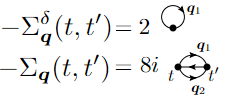}
    \caption{Time-local term and the contribution to the self-energy term due to interaction with reservoir. Here the filled vertex corresponds to the factor $\left(-g_0\right)$, solid lines are reservoir Green's functions $G^{\rm res}_{\bs q}$, given by \eqref{eq:reservoir_green_decomposition}.}
    \label{fig:muSET}
\end{figure}

We may integrate out reservoir degrees of freedom in order to obtain an  effective action for the condensate mode only (with $\psi$ being the corresponding field), which has the following structure:
\begin{align}
        S^{\rm eff}&{=}\int\limits_c d\tau\ovl{\psi}\left\{i\partial_{\tau}{-}\varepsilon_{0}{+}i\frac{\Gamma}{2}{-}\Sigma_0^{\delta}{-}g_0|\psi|^2\right\}\psi\nonumber\\
        &{-}\int\limits_c d\tau\int\limits_c d\tau'\ovl\psi(\tau)\Sigma_0(\tau,\tau')\psi(\tau').
        \label{eq:condensate_effective_action}
\end{align}
Here we introduced a self-energy term $\Sigma$ to describe interaction with reservoir, extracting the time-local contribution $\Sigma^{\delta}(\tau, \tau')=\Sigma^{\delta}\delta(\tau, \tau')$ explicitly. We use the lowest order diagrammatic expressions for these terms, as presented in the Fig. \ref{fig:muSET}.

In order to deal with real-time dynamics, we pass from the field $\psi$ defined on the Keldysh contour (denoted by $c$ in \eqref{eq:condensate_effective_action}) to $\psi_{\pm}$ being the fields on its backward and forward branches. This is achieved by the standard Keldysh rotation
\begin{align}
    \psi_{\pm}=\phi\pm\frac{\xi}{2}.
\end{align}
This procedure makes the real-time action to acquire the following form:
\begin{widetext}
\begin{align}\label{eq:Keldrotated_action}
    &S_{\rm eff}[\phi, \ovl{\phi}, \xi, \ovl{\xi}]=\int_{\tau_0}^{t} d\tau\int_{\tau_0}^{t} d\tau\ovl{\phi}\left[\left\{i\partial_{\tau}{-}\varepsilon{+}i\frac{\Gamma}{2}{-}\Sigma_0^{\delta}{-}g_0|\phi|^2\right\}\delta(\tau,\tau')-\Sigma_0^-(\tau,\tau')\theta(t'{-}t)\right]{\xi'}\nonumber\\
    +&\int_{\tau_0}^{t} d\tau\int_{\tau_0}^{t} d\tau'\ovl{\xi}\left[\left\{i \partial_{\tau}{-}\varepsilon{+}i\frac{\Gamma}{2}{-}\Sigma_0^{\delta}{-}g_0|\phi|^2\right\}\delta(\tau,\tau')-\Sigma_0^+(\tau,\tau')\theta(t{-}t')\right]{\phi'}{-}\frac12\int_{\tau_0}^{t} d\tau\int_{\tau_0}^{t} d\tau'\ovl\xi\Sigma_0^{K}(\tau,\tau')\xi'
\end{align}
\end{widetext}
with local ($\Sigma_0^{\delta}$), retarded ($\Sigma^{+}$), advanced ($\Sigma_0^-$) and Keldysh ($\Sigma_0^{K}$) components of the self-energy term introduced.

To derive the dynamical equation from this type of action, one needs to treat noise terms coupled to the $\phi$ field dynamics, which leads to a Langevin type equation. In the current paper, we focus on the mean-field dynamics by seeking the stationary phase “classical”\ solution in a form:
\begin{align}
    \frac{\delta S_{\rm eff}}{\delta \phi} = 0 &\rightarrow \xi(t) = 0,\\
    \frac{\delta S_{\rm eff}}{\delta \xi} = 0 &\rightarrow \phi(t) = \phi_c(t),
\end{align}
with $\phi_c(t)$ being a solution of the following equation:
\begin{align}\label{eq:statphase_equation}
        i\dot\phi_c{=}\left[\varepsilon{-}i\frac{\Gamma}{2}{+}\Sigma_0^\delta{+}g_0|\phi_c|^2\right]\phi_c(t){+}\int\limits_{t_0}^td\tau \Sigma_0^+(t, \tau)\phi_c(\tau).
\end{align}
Analysing this equation is the main objective of the current work.

Neglecting the noise term imposes several limitations. Namely, we are not able to describe correctly the initial stages of condensate evolution when fluctuations dominate the dynamics. The approach presented below leads to relevant results in the vicinity of stationary points, where noise terms are less significant. Incorporating them terms into the theory is left for future investigations. 
\subsection{Self-energy term}
As we see in the Fig. \ref{fig:muSET}, the self-energy term besides an evident blueshift contribution
\begin{align}
    \Sigma_{\bs q}^{\delta}=g_0\sum_{\bs q_1} f(\zeta_{\bs q_1}{-}\mu)=2g_0N^{\rm res}=V_0 n^{\rm res}
\end{align}
(with $N^{\rm res}$ and $n^{\rm res}$ being the total reservoir occupation and its surface density correspondingly) has a term due to interparticle interaction (here $\tau = t-t'$):
\begin{align}\label{eq:SET_exex}
    \Sigma_{\bs q}(\tau){=}{-}8g_0^2 \sum_{{\bs q_1}, {\bs q_2}} G^{\rm res}_{\bs{q}_{1}}\left(\tau\right) G^{\rm res}_{\bs{q}_{1}+\bs{q}_{2}-\bs q}\left(-\tau\right) G^{\rm res}_{\mathbf{q}_{2}}\left(\tau\right).
\end{align}

Note that the term $\Sigma^{\delta}_{\bs q}$ is independent of $\bs q$, therefore reservoir states are shifted by the same amount and the energy offset $\varepsilon$ is not affected by $\Sigma^{\delta}$. In fact, by this we use the Hartree-Fock approximation for the reservoir (for contact interaction direct and exchange terms give the same contribution, that's why we have a factor of 2 on the first line in the Fig. \ref{fig:muSET}).

For the retarded component, one obtains the following expression by setting $\bs q=0$ (partly following \cite{Elistratov2018} by two of us):
\begin{widetext}
\begin{align}
&\Sigma_0^{+}(\tau)=-8i g_0^{2}e^{-3\gamma_{\rm res} \tau} \sum_{\bs q_1, \bs q_2}e^{-i(\varepsilon_{\bs q_1}+\varepsilon_{\bs q_2}-\varepsilon_{\bs q_1+\bs q_2})t}\left[f_{\boldsymbol{q}_{1}}\left(1+f_{\boldsymbol{q}_{1}+\boldsymbol{q}_{2}}\right) f_{\boldsymbol{q}_{2}}-(1+f_{\boldsymbol{q}_{1}})f_{\boldsymbol{q}_{1}+\boldsymbol{q}_{2}} (1+f_{\boldsymbol{q}_{2}})\right]\Theta(\tau).
\end{align}
\end{widetext}
Considering $\gamma_{\rm res}\to0$ as well as taking advantage of the Bose-Einstein distribution property $1+f(\omega)=e^{\omega}f(\omega)$, we obtain the following expression for the imaginary part of the retarded component:
\begin{align}
    \Im [\Sigma_0^+(\omega)]=&{-}8\pi g_0^2\left[e^{\beta\left(\omega-\mu\right)}{-}1\right]\nonumber\\
    \times\sum_{\bs q_1, \bs q_2}&f_{\bs q_1}(1{+}f_{\bs q_1+\bs q_2})f_{\bs q_2}\delta(\omega+\varepsilon_{\bs q_1+\bs q_2}-\varepsilon_{\bs q_1}-\varepsilon_{\bs q_2}).
    \label{eq:simplified_imret}
\end{align}
Above $f_{\bs q}=f\left[\beta(\varepsilon_{\bs q}-\mu)\right]$.

A dimensionless function $I(\omega)$ may be isolated as follows:
\begin{align}
    \Im [\Sigma_0^+(\omega)]{=}&\frac{V_0^2}{(2\pi)^3\lambda^4_{dB}}I\left(\omega\right)
\end{align}
with $\lambda_{dB}=\frac{\hbar}{\sqrt{2mT}}$ being the thermal de-Broglie wavelength. 
Hereafter $\beta=1$, which means all the energies are measured in units of $kT$.

\begin{figure}[H]
    \centering
    \includegraphics{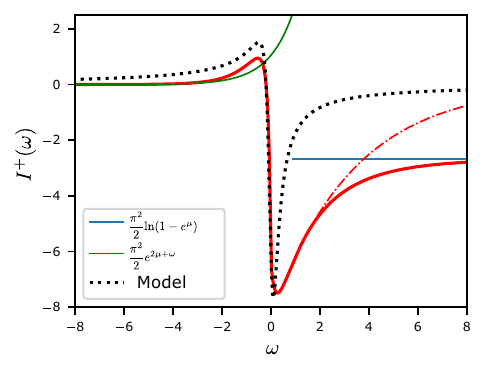}
    \caption{Frequency dependence of the retarded self-energy term component $I(\omega)$ (red solid line) with asymptotics given. The dash-dotted line qualitatively demonstrates how the curve will be modified for finite-range interaction. Here for illustration $\mu = -0.2$ is chosen.}
    \label{fig:I+}
\end{figure}

One may evaluate $I(\omega)$ numerically, considering the quadratic dispersion relation. Moreover, asymptotic behaviour may be studied analytically (see Appendix A  to find calculations for arbitrary momentum $\bs q$). All the information is summarized in the Fig. \ref{fig:I+}.

When discussing the $\omega\to\infty$ asymptotics, we note that $I(\omega)={\rm const}$ does not vanish. This is due to the contribution of the following process: a virtual particle with $q=0$ and $\varepsilon = w$ scatters on a reservoir particle with $q\approx 0,\ \epsilon_{\bs q}\approx 0$. As a result, they are both in the reservoir with  momenta $\bs q_1=-\bs q_2$ and energies $\varepsilon_{\bs q_1}=\varepsilon_{\bs q_2}=\frac{\omega}{2}$. The amplitude of this process does not decrease with growing $\omega$. This is a consequence of the contact interaction model. If we consider some finite interaction radius, \textit{i.e.,} introduce a transfer momentum cut-off, $I(\omega)$ will tend to zero with growing $\omega$ because of growing transferred momentum $q\sim \sqrt{\omega}$ during the scattering process. This is schematically demonstrated by a dash-dotted red line in the Fig. \ref{fig:I+}. The exact behaviour depends, of course, on the interparticle interaction potential.

Using this numerical result for further calculations is quite involved. We will further use an approximation
    \begin{align}
	\Im [\Sigma_0^+(\omega)]{=}&{-}\Lambda^+\frac{\omega-\mu}{\gamma^2+\omega^2},\label{eq:fitting}
\end{align}
which captures the double-peaked shape of the curve as well as reproduces the $x$-intercept correctly, which appears to be important to describe condensate effective chemical potential equilibrating. Here, $\Lambda^+$ and $\gamma$ should be treated as fitting parameters.

Of course, the model curve cannot reproduce the real one exactly. As it will be clear from Sec. \ref{sec:discussion}, when searching the best fit parameters, it's worth better fitting the left peak at the cost of not reproducing the high-frequency behaviour as demonstrated in the Fig. \ref{fig:I+}.

The retarded self-energy term is analytic in the upper half plane in frequency domain due to causality, therefore its real and imaginary part are related by the Kramers-Kronig relations, and we may seek an approximation for the function using its imaginary part only. Therefore:
\begin{align}
    \Sigma_0^+(\omega)=-\Lambda^+\frac{\gamma^2+\omega\mu}{\gamma^2+\omega^2}-i\Lambda^+\frac{\omega-\mu}{\gamma^2+\omega^2}.
\end{align}

In time domain, this function is as follows (note that time is measured in units of $1/kT$):
\begin{align}
	\Sigma_0^+(t)=-\Lambda^+\left(1-i\frac{\mu}{\gamma}\right)e^{-\gamma t}\Theta(t).
 \label{eq:sigma_approx}
\end{align}

\section{Mean-field phase diagram}\label{sec:phdiag}
\subsection{Reduction to an ODE system}

Using \eqref{eq:sigma_approx}, we may write down the dimensionless dynamical equation as follows:

\begin{align}
	i\partial_t\phi_c{=}\Bigg(\varepsilon&{+}g|\phi_c|^2{-}i\frac{\Gamma}2\Bigg)\phi_c \nonumber\\
	&{-}\Lambda^+\left(1{-}i\frac{\mu}{\gamma}\right)\int_0^te^{{-}\gamma(t{-}t')}\phi_c(t')dt'
\label{eq:main_dynequation}
\end{align}

The advantage provided by using an exponential kernel is the possibility to simplify the dynamical equation \eqref{eq:main_dynequation}. To do that, one may consider it along with its time derivative in order to exclude the memory term. Namely, the Madelung transformation $\phi_c(t)=\sqrt{\rho(t)}e^{-i\theta(t)}$ is performed (time arguments are omitted, $\rho'\equiv\rho(t')$, $\theta'\equiv\theta(t')$):
\begin{widetext}
\begin{align}\label{eq:IDE_integral_terms_excluded}
	\partial_t\rho &{=} {-}\Gamma \rho{-}2\sqrt{\rho}\Lambda^+\int_0^t d t' e^{{-}\gamma(t-t')}\sqrt{\rho'}\sin\left[\theta{-}\theta'\right]{+}2\frac{\mu}{\gamma}\sqrt{\rho}\Lambda^+\int_0^t d t' e^{{-}\gamma(t-t')}\sqrt{\rho'}\cos\left[\theta{-}\theta'\right],\nonumber\\
	\rho\partial_t \theta&{=}\left(\varepsilon{+}g\rho\right)\rho{-}\sqrt{\rho}\Lambda^+\int_0^t d t' e^{{-}\gamma(t-t')}\sqrt{\rho'}\cos\left[\theta{-}\theta'\right]{-}\frac{\mu}{\gamma}\sqrt{\rho}\Lambda^+\int_0^t d t' e^{{-}\gamma(t-t')}\sqrt{\rho'}\sin\left[\theta{-}\theta'\right],
\end{align}
\end{widetext}
which leads to the ODE system after differentiating and excluding integral terms:
\begin{align}\label{eq:statphase_ODE}
    \ddot\rho = -\Gamma \dot\rho+\left(\frac{\dot \rho}{2\rho}-\gamma\right)&\left(\dot\rho+\Gamma\rho\right)\nonumber\\
    &+2\frac{\Lambda^+\mu}{\gamma}\rho+2\rho\dot\theta\left(\dot\theta-\varepsilon-g\rho\right),
\end{align}
\begin{align}
	\dot\rho\dot\theta+\rho\ddot\theta=\left(\varepsilon+2g\rho\right)&\dot\rho+\left(\frac{\dot\rho}{2}-\gamma\rho\right)\nonumber\\
	\times&\left(\dot\theta-\varepsilon-g\rho\right)-\frac{\dot\theta}2(\dot\rho+\Gamma\rho)-\Lambda^{+}.
\end{align}
To get rid of negative power terms in $\rho$, we may now use $\rho(t)=e^{r(t)}$ substitution in order to obtain the following autonomous ODE system:
\begin{align}\label{eq:classical_ODEs}
    \ddot r & {=} {-}\left(\frac{\Gamma}2{+}\gamma\right) \dot r{-}\frac{\dot r^2}{2}{+}\left(2\frac{\Lambda^+\mu}{\gamma}{-}\gamma\Gamma\right){+}2\dot\theta\left(\dot\theta{-}\varepsilon{-}ge^r\right),\nonumber\\
	\ddot\theta&{=}\left(\varepsilon{+}2g e^r\right)\dot r{-}\frac32\dot r\dot\theta{+}\left(\frac{\dot r}{2}{-}\gamma\right)\left(\dot\theta{-}\varepsilon{-}ge^r\right){-}\frac{\Gamma\dot\theta}2{-}\Lambda^+{.}
\end{align}
By introducing $V=\dot r$ and $\nu = \dot \theta$ as new variables, we may formulate the dynamical equation as a first order ODE system:
\begin{align}\label{eq:first_order_ODE}
    \dot V &{=} \left(2\frac{\Lambda^+\mu}{\gamma}{-}\gamma\Gamma\right){+}\left(\frac{\Gamma}2{+}\gamma\right) V{-}\frac{V^2}{2}{+}2\nu\left(\nu{-}\varepsilon{-}ge^r\right)\!,\nonumber\\
    \dot \nu&{=}\left(\varepsilon{+}2g e^r\right)V\!{-}\frac{3V\nu}2{+}\!\left(\frac{V}{2}{-}\gamma\right)\!\left(\nu{-}\varepsilon{-}ge^r\right)\!{-}\frac{\Gamma\nu}2{-}\Lambda^+\!\!,\nonumber\\
    \dot r &{=} V.
\end{align}
To derive initial conditions, one may set $t=0$ in \eqref{eq:IDE_integral_terms_excluded}:
\begin{equation}
\begin{cases}
    \partial_t \rho(0) = -\Gamma \rho,\\
    \rho \partial_t\theta (0)=(\varepsilon+g\rho)\rho.
\end{cases}\label{eq:initial_cond}
\end{equation}
This leads to $\nu(0)=\epsilon+ge^r$ and $V(0)=-\Gamma$. Note that these expressions are due to assuming the condensate and reservoir being in contact from $t=0$ exactly. In real system the initial evolution stages may be more complicated, which, however, does not change the asymptotic behaviour of the system.

\subsection{Stationary points}
Here, we seek stationary points of the ODE system \eqref{eq:first_order_ODE} by setting $\dot V = \dot r = \dot\nu  = 0$:
\begin{equation}
\begin{cases}	
	2\nu(\nu - \varepsilon - g\rho)=-2\frac{\mu}{\gamma}\Lambda^++\gamma\Gamma,\\
	\gamma(\nu - \varepsilon - g\rho)+\frac12\Gamma\nu=-\Lambda^+.
\end{cases}
\label{eq:stationary}
\end{equation}
We readily solve these equations to obtain:
\begin{align}\label{eq:satationary_lower_upper}
	\nu_{\pm}&=\frac{-\Lambda^+\pm \sqrt{D}}{\Gamma },\\
	\rho_{\pm}&=\frac1g\left[-\varepsilon{+}\Lambda^+\left(\frac1{2\gamma}{-}\frac{1}{\Gamma}\right)\pm\left(\frac1{\Gamma}{+}\frac{1}{2\gamma}\right)\sqrt{D}\right]\nonumber
\end{align}
with
\begin{align}
	 D=-\gamma^2\Gamma^2+\left(\Lambda^+\right)^2+2\Gamma\Lambda^+\mu.
  \label{eq:discriminant}
\end{align}

For stability analysis of these stationary points, one may consider the linear expansion and obtain the corresponding eigenvalues (see Appendix B). The “upper”\ solution with “$+$" sign appears to be stable and the “lower” one -- unstable.  
\subsection{Decaying solution}
Dealing with a constrained quantity $\rho\geq0$, we should consider one more equilibrating scenario with $\rho\to 0$. We may seek the decaying solution in a form $\ddot r = 0$, $\dot r = \kappa$, $\dot \theta = \Omega = const$. With small $\rho$ being neglected, this leads to the following characteristic equations:
\begin{equation}	
	\begin{cases}
		\frac{\kappa^2}{2}=-\left(\frac{\Gamma}{2}+\gamma\right)\kappa+\left(2\frac{\mu}{\gamma}\Lambda^+-\gamma\Gamma\right)+2\Omega(\Omega-\varepsilon)\\
		0=-\frac{3}{2}\kappa\Omega+\varepsilon\kappa+\left(\frac{\kappa}{2}-\gamma\right)(\Omega-\varepsilon)-\frac{1}{2}\Gamma\Omega-\Lambda^+
	\end{cases}
	\label{eq:decay_stability}
\end{equation}
As derived in Appendix B, there are two eigenmodes with $\kappa<0$ (\textit{i.e.,} the decaying solution is stable) whenever $\rho_+\rho_-\geq0$ or $D<0$. Stated otherwise, the decay solution is stable if there are either no stationary points or both of them are present simultaneously.

\subsection{Phase diagram}
From the discussion above, we infer that there are two possible equilibrating scenarios. The one is reaching a stationary solution $\rho_+$ and the other is a decaying solution. In physical terms, the first one corresponds to condensate formation (with non-zero $\phi$ being the corresponding order parameter) and the second one describes the normal phase.

\begin{figure}
    \begin{center}
    \includegraphics[width=\linewidth]{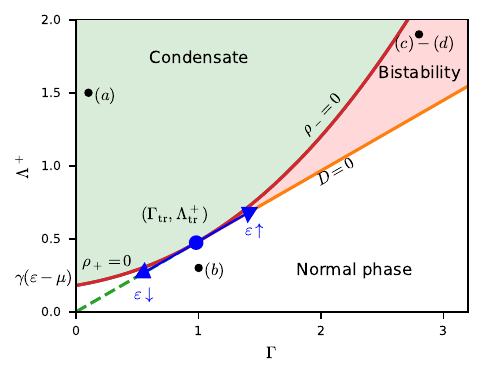}
    \end{center}
    \caption{Phase diagram for the system under consideration. The blue dot denotes the “triple” point of coexistence of all the three phases. The arrows indicate how does this point move with varying $\varepsilon$. Here for demonstration $\gamma = 0.2, g=1, \mu = -0.2, \varepsilon = 0.7$. For each of the black dots (a)-(d) there is an evolution graph presented in the Fig. \ref{fig:nurho}.}
	\label{fig:model_phase_diagram}
\end{figure}

Using the results of the stability analysis, we may summarize them on a phase diagram presented in the Fig. \ref{fig:model_phase_diagram} on ($\Lambda^+, \Gamma$) plane ($\Lambda^+$ physically corresponds to condensate-reservoir interaction “strength”, $\Gamma$ is the decay intensity of condensate particles). The remaining parameters $\gamma$ and $\mu$ are fixed.

On this figure, the condensate exists whenever $\rho_+>0$ (which implies $D>0$). The more strict condition of these two ($\rho_+>0$ and $D>0$) defines the condensate stability boundary.

The decaying solution is stable in the three cases listed below:
\begin{enumerate}
    \item For regions with $D<0$. Here no stationary points exists, decay is the only asymptotic scenario. This is the “Normal phase” region of the diagram, below the straight line.
    \item When $D>0$ but $\rho_-<0$ as well as $\rho_+<0$. There are also no physically relevant stationary points here, this is another part of the “Normal phase” region, which is the in the left bottom corner of the diagram, above the dashed line. Note that for low enough, $\varepsilon$ this region disappears (\textit{e.g.} in the Fig. \ref{fig:oscillations_indicated}).
    \item For $D>0$, $\rho_+>0$ and $\rho_->0$. Here, both the decaying solution and one of the stationary points are stable. This is the "Bistability" region on the phase diagram bounded by $D=0$ and $\rho_-=0$ lines (note that $\rho_-<\rho_+$, that's why it is $\rho_-<0$ which makes the stability criterion $\rho_-\rho_+>0$ invalid). This is kind of an overlap of "Normal phase" and "Condensate" regions of the phase diagram.
\end{enumerate}

The equations of the phase boundaries are presented near the corresponding lines.

There is kind of a triple point on the phase diagram where all the three solutions coexist with $\rho_{\pm}\to 0$. Its position is given by
\begin{equation}
    \begin{cases}
        \Gamma_{\rm tr}=2\gamma\left(1+\frac{\varepsilon}{\sqrt{\gamma^2+\mu^2}-\mu}\right),\\
        \Lambda^+_{\rm tr}=2\gamma\left(\sqrt{\gamma^2+\mu^2}-\mu+\varepsilon\right)
    \end{cases}
\end{equation}
and regardless of the detuning, $\varepsilon$ it is located on the straight line $\Lambda^+=\left(\sqrt{\gamma^2+\mu^2}-\mu\right)\Gamma$ (as indicated by blue arrows on the dashed line in the Fig. \ref{fig:model_phase_diagram}).

Note that for $\Gamma\to 0$, only the “upper” stationary point exists. It is given by:
\begin{align}
\nu&=\mu,\\
\rho&= \frac1g\left[\mu-\varepsilon+\frac{\Lambda^+}{\gamma}\right].
\end{align}
The effective chemical potential $\nu$ of the condensate becomes equal to the one of the particle reservoir, as one could expect for atomic gas of long-living particles. The condensation threshold may be identified at $\rho\to 0$ (see the $y$-intercept in the Fig. \ref{fig:model_phase_diagram}):
\begin{align}
    \Lambda^+ = \gamma(\varepsilon - \mu).
\end{align}
Since $\Lambda^+, \mu$ and $\gamma$ themselves are not independent quantities, but they depend on density and temperature, this equation may be treated as a one defining the critical effective temperature $T_c^{\rm eff}$. When increasing condensate decay rate, this temperature gets shifted. A demonstration of critical temperature evaluation will be presented in Sec. \ref{sec:demonstration} where we map this phase diagram on the density/temperature plane.

\section{Dynamics}\label{sec:dyneffects}

\subsection{Evolution in different regimes}

\begin{figure}
    \centering
    \includegraphics{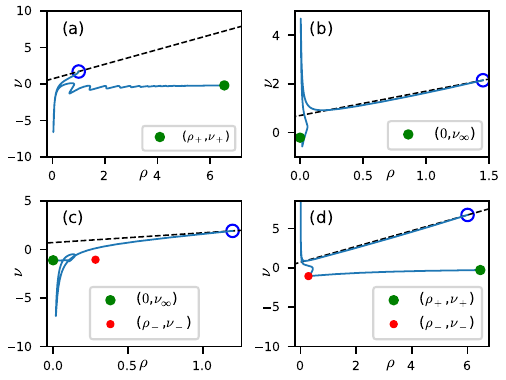}
    \caption{Evolution illustrated for (a) “Condensate” phase; (b) “Normal phase”; (c)-(d) “Bistability” phase for two initial occupations. The letters correspond to points in the Fig. \ref{fig:model_phase_diagram}. The dashed line restricts the initial conditions by $\nu=\varepsilon+g\rho$ as given by \eqref{eq:initial_cond}. The hollow circle indicates the initial occupation chosen.}
    \label{fig:nurho}
\end{figure}

In the two of the three phases described, condensate formation is possible. In the “Condensate” phase there is a single stationary point present which attracts all the ODE solutions regardless of the initial conditions as demonstrated in the Fig. \ref{fig:nurho} (a).

In contrast, in the “Normal phase” region all the solutions are attracted towards $\rho=0$ with $\dot \theta$ approaching $\nu_{\infty}$, which corresponds to the slower decaying eigenmode as illustrated in Fig \ref{fig:nurho} (b) (see Appendix B for details).

In the bistability regime, an unstable stationary point appears which repels the occupation to either $\rho = 0$ or the stable point as illustrated in the Fig. \ref{fig:nurho} (c)-(d). Note that in the Fig. \ref{fig:nurho} (c) the upper stationary point also exists at a higher occupation.

One may infer, studying the evolution in the bistable regime, that higher initial occupations are attracted to the stationary point (condensate formation takes place as in the Fig. \ref{fig:nurho} (d)) and lower ones decay to zero as presented in the Fig. \ref{fig:nurho} (c). However, it is not always the case, since for some parameters even for large $\rho$ the line of initial condition $\nu = \varepsilon + g\rho$ does not intersect the attraction basin of the stable point.

\subsection{Condensate formation. Relaxation oscillations}
On fig \ref{fig:nurho} (a) one may see oscillations when approaching equilibrium. Though, such type of relaxation oscillations are not a general feature of the system.

As it is shown in details in Appendix B, the eigenmode expansion of $\phi(t)$ close to stationary state consists of terms $e^{\kappa_0 t}$, $e^{(\kappa_1\pm i\Omega) t}$. We expect significant asymptotic density oscillations similar to the ones described in the scope of a different model in \cite{Opala2018}, when $\kappa_0<\kappa_1$ (note that both are negative) in order for the oscillatory terms to dominate at late times. This regime is illustrated in the Fig. \ref{fig:oscidemo} (a)-(c). In contrast, for $\kappa_0>\kappa_1$, at late times condensate occupation monotonously approaches stationary value as shown in the Fig. \ref{fig:oscidemo} (d)-(f).

\begin{figure}[h]
    \centering
    \includegraphics{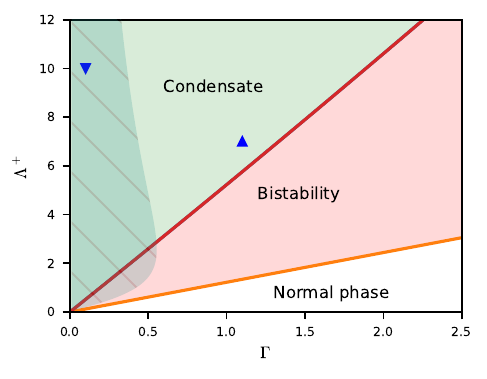}
    \caption{Phase diagram for low enough $\varepsilon$. In the hatched region $\kappa_0<\kappa_1$, asymptotic dynamics is oscillatory. Here for demonstration $\gamma = 1,\ g=1,\ \mu = -0.2,\ \varepsilon = -10$.}
    \label{fig:oscillations_indicated}
\end{figure}

\begin{figure}[h]
    \centering
    \includegraphics{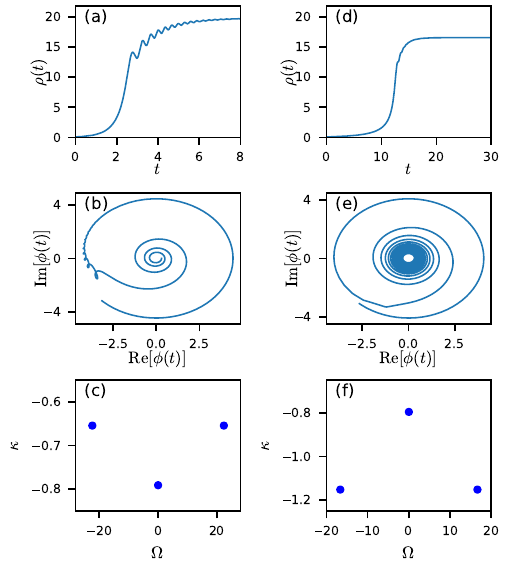}
    \caption{(a), (b) Relaxation oscillations at $\Gamma = 0.1$, $\Lambda^+ = 10$. (c) Eigenmodes for this case. These plots correspond to {\color{blue} $\blacktriangledown$} in the Fig. \ref{fig:oscillations_indicated}.(d), (e) Monotonous asymptotics for $\Gamma = 1.1$, $\Lambda^+ = 7$. (f) Eigenmodes for this case. These plots correspond to {\color{blue} $\blacktriangle$} in the Fig. \ref{fig:oscillations_indicated}.}
    \label{fig:oscidemo}
\end{figure}

Relaxation oscillations are present for low enough $\varepsilon$, the typical phase diagram for this regime is presented in the Fig. \ref{fig:oscillations_indicated}.

Of course, the oscillations may be observable outside the hatched region of the phase diagram also. However, at late times, they are replaced by monotonous exponential approach to the stationary state.

Note that oscillations are damped for higher values of $\Gamma$, which may be considered as a consequence of increasing the impact of memory terms by means of decreasing the lifetime of the particles in the condensate.

\subsection{Bifurcation scheme}
\begin{figure}[h]
\begin{tikzpicture}[scale=1]
            \begin{axis}[
                name = negative,
                axis x line = center,
                axis y line = center,
	        xmin=0, xmax=6, ymin=-0.5, ymax=11,
	        xtick=\empty,
	        ytick=\empty,
	        height=5cm, width=9cm, grid=none,
	        xlabel={$\Lambda^+$},
	        ylabel={$\rho$},
	        samples = 100
	]
    \begin{scope}[>=stealth]
	\addplot[mark=none, domain = 0:4.174, ultra thick, blue, samples=300, dashed] {-((0.5))+(1.72414*((-0.58+(3))*x-(0.58+(3))*sqrt(-0.0841*(3)^2-0.8*(3)*x+x^2)))/(3)};
	\addplot[mark=none, domain = 0:2.5, violet!70, dashed, arrow inside={end={<},opt={violet!70, scale=2.1}}{0.5}]  coordinates{(2.686,0)(2.686,3.23)};
	\addplot[mark=none, domain = 0:2.5, violet!70, dashed, arrow inside={end={<},opt={violet!70, scale=2.1}}{0.7}]  coordinates{(4.174,10.7)(4.174,0)};
	\addplot[mark=none, domain = 0:6, ultra thick, blue, samples=300, arrow inside={end={<},opt={violet!70, scale=1.3}}{0.25}] {-((0.5))+(1.72414*((-0.58+(3))*x+(0.58+(3))*sqrt(-0.0841*(3)^2-0.8*(3)*x+x^2)))/(3)};
	\addplot[mark=none, domain = 0:4.174, ultra thick, blue, arrow inside={end={>},opt={violet!70, scale=1.3}}{0.85}] {0.005};
	\end{scope}
	\addplot[mark=none, domain = 0.55:6, dashed, black] {-0.5+(1/0.58-1/3)*x};
	\draw [thick,-{Stealth[slant=0]}] (axis cs: 2.686,3.23) -- (axis cs: 2.686+0.7, 3.23+1.39*0.7);
	\draw [thick,-{Stealth[slant=0]}] (axis cs: 2.686,3.23) -- (axis cs: 2.686-0.7, 3.23-1.39*0.7);
	\node[] at (axis  cs:2.686+0.6, 3.23+1.39*0.25-0.45) {$\Gamma\bs\uparrow$};
	\node[] at (axis  cs:2.686-0.6, 3.23-1.39*0.25+0.4) {$\Gamma\bs\downarrow$};
	\draw[black, fill=white] (2.686,3.23) circle (2pt);
	\path [name path=xaxis]
      (\pgfkeysvalueof{/pgfplots/xmin},0) --
      (\pgfkeysvalueof{/pgfplots/xmax},0);
	\path [name path=xup]
      (\pgfkeysvalueof{/pgfplots/xmin},11) --
      (\pgfkeysvalueof{/pgfplots/xmax},11);
	\addplot[red!50, opacity=0.4] fill between [of=xaxis and xup, soft clip={domain=2.686:4.174}];
    \definecolor{dark green}{rgb}{0,0.5,0}
	\addplot[dark green!50, opacity=0.4] fill between [of=xaxis and xup, soft clip={domain=4.174:6}];
	\draw [-{Stealth[slant=0]}] (axis cs: 5,1.39*5) -- (axis cs: 5, -0.5+1.39*5+0.9);
	\draw [-{Stealth[slant=0]}] (axis cs: 5,1.39*5) -- (axis cs: 5, -0.5+1.39*5-0.9);
	\node[] at (axis  cs: 5+0.25, -0.5+1.39*5-0.9) {$\varepsilon\bs\uparrow$};
	\node[] at (axis  cs: 5-0.25, -0.5+1.39*5+0.9) {$\varepsilon\bs\downarrow$};
\end{axis}
	\filldraw[blue!80] (1.05-0.5,0.15) circle (3pt);
	\node[below] at  (1.05-0.5,0) {$\Lambda^+_{tr}$};
 \end{tikzpicture}
 \caption{Bifurcation diagram for the bistable region of the phase diagram. The stable solution branches are drawn as solid curves, for the unstable one a thick dashed curve is used. The bifurcation point is pictured in white. Arrows indicate its movement along the dashed line with $\Gamma$ being changed. The double arrow on the right indicates how does the line move with $\varepsilon$ being varied.}
 \label{fig:bifurcation_diagram}
\end{figure}
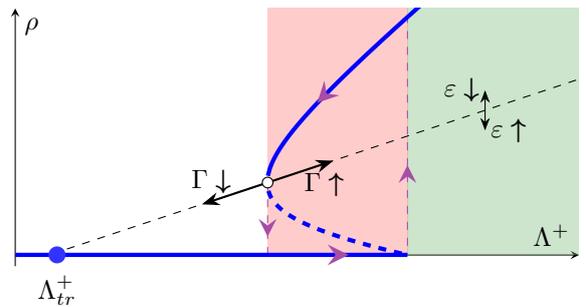

In order to draw several physical predictions in the bistable region, we study here the bifurcation diagram of the ODE system. It is presented schematically in the Fig. \ref{fig:bifurcation_diagram} for fixed $\Gamma$, $\mu$, $g$ and $\varepsilon$. It may be treated as a cross-section of the phase diagram Fig. \ref{fig:model_phase_diagram} by a vertical line passing to the right of the “triple point”.

We readily observe the bistable region where there are two stable branches. Along with the repelling behaviour of the lower stationary point, illustrated in the Fig. \ref{fig:nurho} (c)-(d), we expect two physical effects in this region.

Foremost, when condensate is formed in the bistability region, hysteresis is possible when changing the $\Lambda^+$ by varying the quasiparticle density. This is illustrated by a cycle of violet arrows in the Fig. \ref{fig:bifurcation_diagram}.

The second prediction is related to condensate formation dynamics. 
One should note that initial conditions for $\phi_c$ cannot be defined precisely, at least due to uncertainty relation. We may set up the initial distribution only. This makes condensate formation a probabilistic process. In any particular realization, with all the other parameters being the same, the system may end up either in normal or in condensed phase.

Detailed study of both these effects is beyond the scope of the mean-field analysis, fluctuations should be systematically treated.
\section{Discussion}\label{sec:discussion}
\subsection{Alternative treatment. The origin of phases}
The integro-differential equation \eqref{eq:statphase_equation} itself may provide some useful qualitative understanding even without converting it to an ODE system. Namely, considering a stationary solution in a form $\phi_c=\sqrt{\rho} e^{-i\nu t}$ (assuming non-zero $\rho$ and $\nu={\rm const}$), we may derive the following pair of equations:
\begin{equation}
\begin{cases}
	\nu = \left[\varepsilon_0+\Sigma^{\delta}+g\rho\right]+\Re\left[\Sigma_0^+(\nu)\right],\\
	\Im\left[\Sigma_0^+(\nu)\right]-\frac{\Gamma}2=0.
\end{cases}\label{eq:equi_eqsys}
\end{equation}
The first of them provides a relation for the self-consistent effective chemical potential $\nu$ of the condensate. The second one describes particle flux saturation.
Obviously, we should impose a constraint $g\rho\geq0$, therefore, the necessary condition for condensation is existence of a non-empty set of zeros $\nu^*$ of the imaginary part $\Im[\Sigma_0^+(\nu)]$, which satisfy the condition $\Re\left[\Sigma^+(\nu^*)\right]\leq\nu^*-\varepsilon_0-\Sigma^{\delta}$. This condition may be considered as another version of the condensate formation criterion described by H.T.C. Stoof in \cite{Stoof_coher} when describing spontaneous symmetry breaking in cold atomic gas.

These general statements may be illustrated with the use of the model expression for the self-energy term. The equation \eqref{eq:main_dynequation} leads to
\begin{equation}
    \begin{cases}
        \nu=\varepsilon+g\rho-\frac{\Lambda^+}{\gamma}\frac{\gamma^2+\nu\mu}{\gamma^2+\nu^2},\\
        \frac{\Gamma}{2}=\Lambda^+\frac{\mu-\nu}{\gamma^2+\nu^2}.
    \end{cases}\label{eq:equi_eqsys_model}
\end{equation}

\begin{figure}[H]
	\centering
	\begin{tikzpicture}[scale=1]
            \begin{axis}[
            	axis x line = center,
            	axis y line = center,
	            xmin=-4, xmax=2.8, ymin=-2, ymax=1.25,
	            xtick=\empty,
	            ytick=\empty,
	            height=7cm, width=9cm, grid=none,
	            xlabel={$\nu$},
                    legend pos=south west
	        ]
	        \addplot[mark=none, orange, thick, domain=-4:0, samples = 300]{-3.5*(x+0.15)/(4*x*x+1)};
	        \addplot[mark=none, blue, dash dot,ultra thick, domain=-4:0, samples = 300]{-(3.5/2)*(0.25-0.15*x)/(x*x+0.25)};
	        \addplot[name path = thrline, mark = none, red, thick, domain=-4:4, samples= 100]{0.7*x+1};
	        \addplot[mark = none, black, dotted, ultra thick, domain=-4:0, samples= 100]{0.7*x+1-2.35};
	        \legend{$\Im[\Sigma^+(\nu)]$, $\Re[\Sigma^+(\nu)]$, $\nu -\varepsilon$, $\nu-\varepsilon-g\rho_+$};
	        \addplot[mark=none, orange, thick, domain=0:1.5, samples = 300]{-3.5*(x+0.15)/(4*x*x+1)};
	        \addplot[mark=none, blue, dash dot,ultra thick, domain=0:1.5, samples = 300]{-(3.5/2)*(0.25-0.15*x)/(x*x+0.25)};
	        \addplot[mark = none, black, dotted, ultra thick,  domain=0:1.5, samples= 100]{0.7*x+1-2.35};
	        \addplot[mark=none, orange, thick, domain=1.5:4, samples = 300]{-3.5*(x+0.15)/(4*x*x+1)};
	        \addplot[mark=none,  blue, dash dot,ultra thick, domain=1.5:4, samples = 300]{-(3.5/2)*(0.25-0.15*x)/(x*x+0.25)};
	        \addplot[mark = none, black, dotted, ultra thick, domain=1.5:4, samples= 100]{0.7*x+1-2.35};
	        \filldraw[blue!80] (-2.7,0.3) circle (2pt);
	        \filldraw[blue!80] (-0.28,0.3) circle (2pt);
	        \addplot[mark=none, black, thick, dashed, domain=-3.5:0.1]{0.3};
	        \node[] at (axis cs:0.17,0.3) {$\frac{\Gamma}{2}$};
	        \node[] at (axis cs:-0.29,1) {$-\varepsilon$};
	        \node[] at (axis cs: -0.46, 0.1) {$\nu_+$};
	        \node[] at (axis cs: -2.8, 0.1) {$\nu_-$};
	        \addplot[name path=xaxis]{-3};
	        \addplot[dotted] (-2.7, -4) -- (-2.7, 4);
	        \addplot[dotted] (-0.28, -4) -- (-0.28, 4);
	        \addplot[dashed, domain = -0.28:0.72]{1-0.2};
	        \addplot[dashed, domain = -0.28:0.72]{1-2.35-0.2};
	        \draw [-{Stealth[slant=0]}, thick] (axis cs: 1-0.28,-0.175-0.2) -- (axis cs: 1-0.28, 1-0.2);
	        \draw [-{Stealth[slant=0]}, thick] (axis cs: 1-0.28,-0.175-0.2) -- (axis cs: 1-0.28, -1.35-0.2);
	        \node[rotate = 90] at (axis cs: 1.35-0.28-0.15, -0.18-0.2) {$g\rho_+$};
	        \node[] at (axis cs: -2.7-0.15,-0.15-0.21) {$A_-$};
	        \node[] at (axis cs: -0.28-0.27,-1.55+0.03) {$A_+$};
        \addplot[green!10, opacity=0.4] fill between [of=thrline and xaxis, soft clip={domain=-4:4}];
        \filldraw[red!80] (-2.7,-0.15) circle (2pt);
	    \filldraw[red!80] (-0.28,-1.55) circle (2pt);
	        \end{axis}
	\end{tikzpicture}
	\caption{Solving graphically the equation system \eqref{eq:equi_eqsys_model}}
	\label{fig:Imsig}
\end{figure}
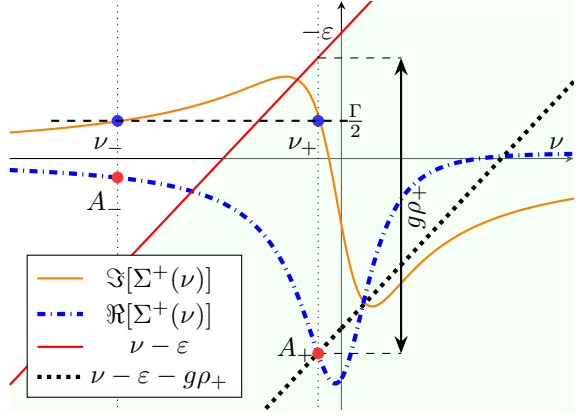

The stationary points $(\nu_\pm, \rho_{\pm} )$ may be obtained from the graphical representation of the equation system \eqref{eq:equi_eqsys_model} below in the Fig. \ref{fig:Imsig}.

One may identify two intersection points of $\Im[\Sigma^+(\nu)]$ with $\frac{\Gamma}{2}$ in the Fig. \ref{fig:Imsig} as the ones, corresponding to the two stationary points \eqref{eq:satationary_lower_upper}. By solving the first equation of the system with $\nu_{\pm}$ with respect to $\rho$, we find stationary occupation numbers $\rho_{\pm}$, which are physically relevant if $\rho\geq0$. Graphically, $\rho_{\pm}>0$ when the corresponding points $A_{\pm}$ with coordinates $(\nu_{\pm}, \Re[\Sigma^+(\nu_{\pm})])$ are in the shaded region below the line $\nu-\varepsilon$. For instance, in the Fig. \ref{fig:Imsig} only one stationary point $A_+$ is in the shaded region, which corresponds to the "Condensate" phase.

This qualitative discussion provides an illustration for the claim from Sec. \ref{sec:model_equations} about focusing on the left peak when fitting the $\Im[\Sigma^+(\omega)]$ curve, since it defines the stationary points. This is by no means a fully justified statement, since the real part of the self energy may be significantly modified even in the vicinity of $\omega=0$ along with changing the high-frequency behaviour of the imaginary part. It means that our model is incapable of describing fine effects due to the exact form of the interparticle interaction (which defines mostly how exactly does the $\Im[\Sigma(\omega)]$ decay at high frequencies). It is suitable for qualitative description only.

However, for an arbitrary $\Sigma^+(\omega)$, one can apply the same graphical procedure and seek solutions in the shaded region where $\nu-\varepsilon\geq\Re[\Sigma(\nu)]$. This approach allows not only equilibrating the occupation $\rho=|\phi|^2$ of the condensate, but also describing the spontaneous symmetry breaking and stationary phase dynamics.
\subsection{Reservoir particle interaction}
As one may infer from equation system \eqref{eq:equi_eqsys}, the condensate formation is crucially dependent on the form of $\Re[\Sigma_0^+(\omega)]$. Moreover, since reservoir levels may be shifted also, the relative offset $\Re[\Sigma_{0}^{+}(\omega)-\Sigma_{\bs q}^{+}(\omega)]$ with respect to reservoir states with non-zero $\bs q$ is relevant. Taking this into account will change the quantitative predictions of the theory. However, as  demonstrated in Appendix A, the asymptotic behaviour of $\Sigma^+(\bs q)$ is independent of $q$, the double-peaked shape remains the same as well as $\Sigma_{\bs q}^+(\mu)=0$ for arbitrary $\bs q$. That's why we hope that a thorough treatment (the one similar to what is done for 3D cold atom gas in Ref. ~\cite{Stoof_instability_9192}) will not to affect qualitative predictions of the model.

\subsection{Late time evolution}
We may now show how the model equation  \eqref{eq:main_dynequation} is related to the driven-dissipative Gross-Pitaevski model \eqref{eq:ODGPE} (for uniform system since we deal with a single-level condensate). To do that, one should consider the particle flux dependence on the condensate occupation number. As discussed above, using an ansatz $\phi=\sqrt{\rho}e^{-i\nu t}$, one may derive the equation pair \eqref{eq:equi_eqsys_model} from \eqref{eq:main_dynequation}. By expanding near the stable stationary point $(\nu_+, \rho_+)$, we express the particle flux as follows:
\begin{align}
    d_t|\phi|^2=2(\gamma_{\mathrm{eff}}-\Gamma_{\mathrm{eff}}|\phi|^2).
\end{align}
with the coefficients given by
\begin{align}
    \Gamma_{\mathrm{eff}}&=\frac{2g\rho_+\Lambda^+ \gamma\left(\gamma^2{+}(2\mu{-}\nu_+)\nu_+\right)}{\gamma(\gamma^2{+}\nu_+^2)^2{-}\Lambda^+\nu_+(2\gamma^2{+}\mu\nu_+){+}\gamma^2\Lambda^+\mu},\\
    \gamma_{\mathrm{eff}}&=\Gamma_{\rm eff}\rho_+.
\end{align}
We have here the coefficients of \eqref{eq:ODGPE} expressed in terms of the ones of \eqref{eq:equi_eqsys_model} in the vicinity of the stable point. For long-living particles with $\Gamma=0$ (which leads to $\nu_+=\mu$) the expression for $\Gamma_{\rm eff}$ is given by
\begin{align}
    \Gamma_{\rm eff}= \frac{g\gamma\Lambda^+}{\gamma^3-\Lambda^+\mu+\gamma\mu^2}.
\end{align}

However, note that in contrast to \eqref{eq:ODGPE}, the model approach developed here with an exponential memory kernel describes how the reservoir imposes not only the occupation but the condensate effective chemical potential also. This is due to frequency dependent gain, which is described by the frequency dependence of the memory kernel.

This is crucial for describing condensate formation and its phase dynamics. That's what allows us to identify the phase boundaries.

\subsection{Numerically fitting the memory kernel}
In order to adopt the presented model for describing real-life systems, one needs to perform numerical integration over the polariton momenta in \eqref{eq:simplified_imret} (see Appendix A for details) and then use fitting to evaluate $\Lambda^+$ and $\gamma$. Remarkably, this can be done just once since the dimensionless function $I^+(\omega)$, which was introduced in Sec. \ref{sec:model_equations} is only dependent on the normalized density $\tilde n=n\lambda_{db}^2$. Performing the fitting for various $\tilde n$ (see details at the end of Appendix A) results in approximate expressions of the form:
\begin{align}
    \Lambda^+&=   \begin{cases}
        \frac{16.94}{(2\pi)^3}\tilde n^{1.55} & \text{if } \tilde n \leq 0.18\\
        \frac{4.94}{(2\pi)^3}\tilde n^{0.83} & \text{if } \tilde n > 0.18
    \end{cases},\\
    \gamma &= 0.56 e^{-9.79 \tilde n},
    \label{eq:fitting_approx}
\end{align}
which are reasonably accurate for $\tilde n \in [0.01;1]$. 

\section{Demonstration}\label{sec:demonstration}
Above, we studied the system, given by \eqref{eq:Hamiltonian}. Here we will demonstrate how to adopt the results to a particular quasiparticle system such as an exciton-polariton gas.

Generally, for low enough temperatures, one may consider lower polaritons with dispersion
\begin{align}
    E_{\rm LP}(k){=}&\frac{k^2}4\left[\frac1{m_{\rm ph}}{+}\frac1{m_{\rm ex}}\right]\nonumber\\
    {-}&\frac12\sqrt{\left(\frac{k^2}2\left[\frac1{m_{\rm ph}}{-}\frac1{m_{\rm ex}}\right]{+}\Delta_0\right)^2{+}4\Omega^2}
\end{align}
Here $\Omega$ is the Rabi splitting, $m_{\rm ph}$ and $m_{\rm ex}$ stand for photon and exciton masses respectively, $\Delta$ is the photon dispersion detuning with respect to the excitonic one. It is given as follows ($E_g$is the semiconductor gap, $E_b<0$ is the exciton binding energy):
\begin{align}
    \Delta_0=\frac{hc}{2D}-E_g-E_b
\end{align}
with $D$ being here the microcavity width.
Condensate is mainly localized at the minimum, and the reservoir particles occupy the “flat” part of the spectrum. Therefore, we may argue that the condensate offset is given by
\begin{align}
    \varepsilon=\frac{\Delta_0}{2}-\sqrt{\left(\frac{\Delta_0}{2}\right)^2+\Omega^2},
\end{align}
which is negative regardless of the sign of $\Delta$.

\begin{figure}[H]
	\centering
	\begin{tikzpicture}[scale=1]
            \begin{axis}[
            	axis x line = none,
            	axis y line = center,
	            xmin=-4, xmax=4, ymin=-1.7, ymax=3.7,
	            xtick=\empty,
	            ytick=\empty,
	            height=7cm, width=9cm, grid=none,
	            xlabel={$k$},
	            ylabel={$E(k)$}
	        ]
	        \addplot[mark=none, orange, thick, domain=-4:4, samples = 300]{0.5*((x*x*5.0148)-sqrt((4.98516*x*x-2.6)^2+4.58))};
	        \addplot[mark=none, orange, domain=-4:4, samples = 300]{0.5*((x*x*5.0148)+sqrt((4.98516*x*x-2.6)^2+4.58))};
	        \addplot[mark=none, dashed, domain=-3:3, samples = 300]{0.0148*x^2+1.3};
	        \addplot[mark=none, dashed, domain=-3:3, samples = 300]{5*x^2-1.3};
	        \draw[<->] (1,-1.6537) -- (1,1.27) node at (1.2, 0) [anchor=west] {$|\varepsilon|$};
	        \draw[dashed] (-3,-1.6837)--(3,-1.6837);
	        \draw[-latex] (-2,0) -- (-0.2,-1.6837) node at (-3.9, 0.2) [anchor=west] {Condensate};
	        \draw[-latex] (2.2,1.85) -- (1.9,1.34) node at (1.5, 2.1) [anchor=west] {Reservoir};
	        \end{axis}
	\end{tikzpicture}
	\caption{Polariton dispersion. For the lower polariton branch, the energy detuning $\varepsilon$ of the condensate with respect to the reservoir is indicated.}
\end{figure}
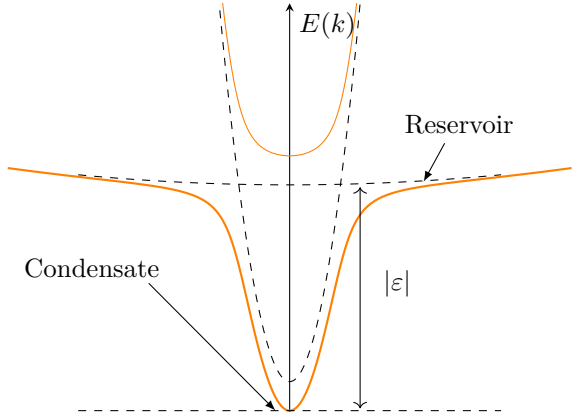

\begin{figure*}[t]
    \includegraphics{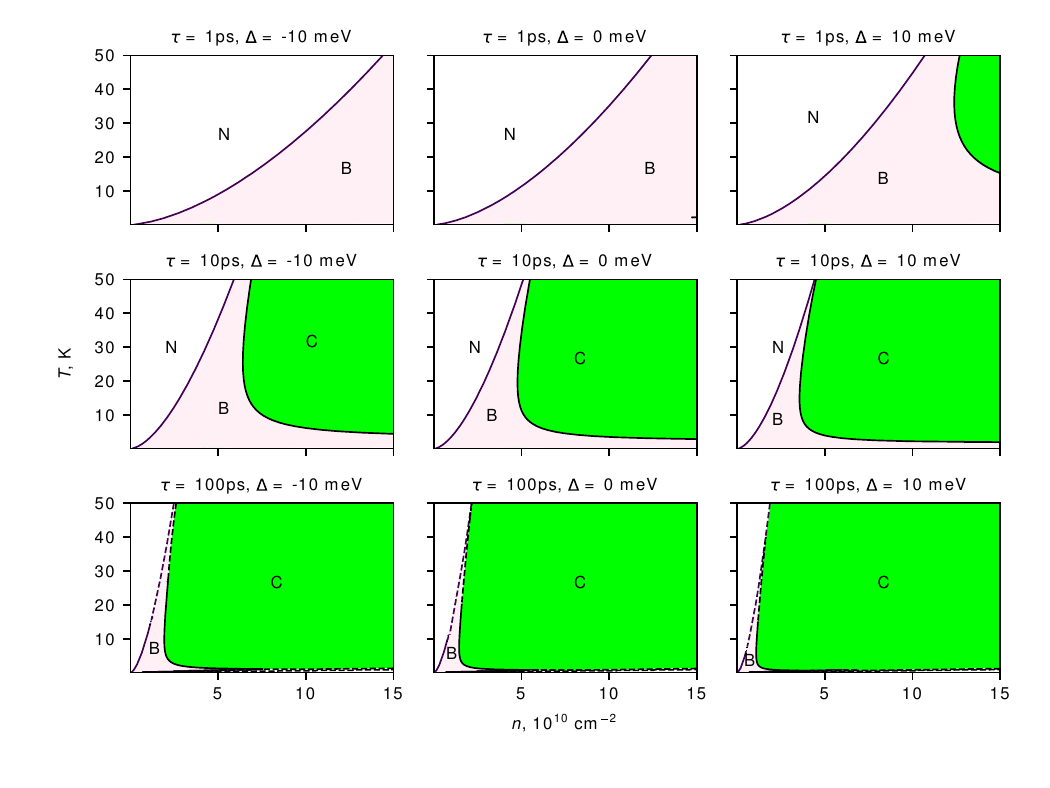}
    \caption{Phase diagrams for CdTe microcavity polaritons for various photon lifetimes $\tau$ and detunings $\Delta$. On the plots "N", "B" and "C" stand for "Normal phase", "Bistability" and "Condensate" respectively. Phase boundaries are dashed in the regions where the fitting procedure and the analytical approximation \eqref{eq:fitting_approx} are no more reliable ($\tilde n$ is outside the region $[0.01;1]$).}
    \label{fig:CdTe_phase_diagram}
\end{figure*}

When describing polariton-polariton interaction, we need to take into account Hopfield coefficients. Namely, the excitonic coefficient is given by ($k$ is the polariton momentum)
\begin{align}
    |X_k|^2=\frac12\left(1+\frac{\Delta_k}{\sqrt{\Delta_k^2+4\Omega^2}}\right)
\end{align}

We further use some simplifications. Namely, given that excitonic mass is usually negligible compared to the photonic one (the ratio is of order $10^{-4}-10^{-3}$), we note that the momentum-dependent detuning $\Delta$ significantly exceeds Rabi splitting $\Omega$ in case    \begin{align}
        \Delta \approx \Delta_0+\frac{k^2}{2m_{\rm ph}}\gg \Omega
    \end{align}
for typical reservoir momenta. Given the reservoir temperature, this assumption is valid for $kT\gg\frac{m_{\rm exc}}{m_{\rm ph}}\Omega$. The typical Rabi splitting for quantum well excitons is of order $\Omega\sim 10$ meV, this assumption is violated only for extremely low temperatures $T\ll 1$ K.

That's why we adopt this assumption ($\Delta\gg \Omega$ for reservoir) to consider the reservoir as dominantly excitonic with $|X_k|^2\approx 1$.

Since interpolariton interaction is mainly due to the excitonic component, we may use a model Hamiltonian of the following form:
    \begin{align}\label{eq:pol_Hamiltonian}
    &\hat H  {=} \left(\varepsilon{-}i\frac{\Gamma}{2}\right)\hat \psi_0^{\dagger}\hat \psi_0{+}\sum_{\bs q\neq\bs 0}\varepsilon_{\bs q}\hat \psi_{\bs q}^{\dagger}\hat \psi_{\bs q}{+}g_0 |X_0|^4 \hat{\psi}_{0}^{\dagger} \hat{\psi}_{0}^{\dagger} \hat{\psi}_{0} \hat{\psi}_{0}\nonumber\\
    &{+}g_0|X_0|^2\! \!\sum_{\boldsymbol{q}^{\prime}{\neq} \bs 0}\!\left[\hat{\psi}_{\bs q^{\prime}}^{\dagger} \hat{\psi}_{{-}\boldsymbol{q}^{\prime}}^{\dagger} \hat{\psi}^2_{0}{+}4\hat{\psi}_{\bs q^{\prime}}^{\dagger} \hat{\psi}_{0}^{\dagger} \hat{\psi}_{0} \hat{\psi}_{\boldsymbol{q}^{\prime}}{+}(\hat{\psi}_{0}^{\dagger})^2\hat{\psi}_{\boldsymbol{q}^{\prime}}  \hat{\psi}_{-\boldsymbol{q}^{\prime}}\!\right]\nonumber\\
    &{+}2g_0|X_0|\!\!\sum_{\boldsymbol{q}_{1}, \boldsymbol{q}_{2}\neq \bs 0} \left[\hat{\psi}_{0}^{\dagger}\hat{\psi}_{\boldsymbol{q}_{1}{+}\bs q_2}^{\dagger} \hat{\psi}_{\boldsymbol{q}_{1}} \hat{\psi}_{\boldsymbol{q}_{2}}{+}\hat{\psi}_{\boldsymbol{q}_{1}}^{\dagger} \hat{\psi}_{\boldsymbol{q}_{2}}^{\dagger}\hat{\psi}_{\boldsymbol{q}_{1}{+}\bs q_2}\hat{\psi}_{0}\right]\nonumber\\
    &{+}g_0 \sum_{\boldsymbol{q}_{1}, \boldsymbol{q}_{2}\neq \bs 0} \hat{\psi}_{\boldsymbol{q}_{1}{+}\bs q^{\prime}}^{\dagger} \hat{\psi}_{\boldsymbol{q}_{2}{-}\boldsymbol{q}^{\prime}}^{\dagger} \hat{\psi}_{\boldsymbol{q}_{1}} \hat{\psi}_{\boldsymbol{q}_{2}}.
\end{align}

This leads to an additional factor of $|X_0|^2$ for $\Sigma^+$. Moreover,  blueshifts for the condensate and the reservoir are different now, and we need to renormalize the offset energy as follows:
\begin{align}
    \tilde\varepsilon = \frac{\Delta}{2}-\sqrt{\left(\frac{\Delta}{2}\right)^2+\Omega^2} {+} \left(|X_0|^2-1\right)V_0n.
\end{align}

When considering condensate polariton decay rate, one may express it as a function of cavity photon lifetime $\tau$:
\begin{align}
    \Gamma=\frac1{\tau}\left(1-|X_0|^2\right)
\end{align}
with $\left(1-|X_0|^2\right)$ representing the photonic component of the condensate polariton.

As a real-life example, we may treat in this fashion a polariton gas in CdTe/CdMgTe microcavites ~\cite{Kasprzak}. The necessary parameters are $\Omega=26$ meV, $m_{\rm exc}=0.69$ meV, $V_0=1.8$ $\mu eV\times \mu m^2$.

With the help of the approach described above, we solve numerically the equations $D=0$, $\rho_{\pm}=0$ (which define the phase boundaries as illustrated in the Fig. \ref{fig:model_phase_diagram}).

In the Fig. \ref{fig:CdTe_phase_diagram} the phase diagram is presented in terms of reservoir density and temperature for several experimentally accessible detuning values and condensate lifetimes. It shows how the lines in the Fig. \ref{fig:model_phase_diagram} are mapped on density-temperature plane and allows localizing all the three phases. These plots demonstrate that the bistable region is indeed more significant for high decay rates of condensate particles (lower lifetimes), its existence is due to intense condensate decay.

\section{Conclusion}

In this paper, we considered a model for describing quasiparticle condensate as an open system embedded in a quasi-equilibrium reservoir. The corresponding open-dissipative Gross-Pitaevski type equation for the condensate has an integral memory term due to the influence of the reservoir. We proposed a simplification, which allows treating the complex integro-differential dynamical equation as an autonomous ODE system. Dealing with stationary solutions of this system, we described a phase diagram predicting the existence of a bistable phase. Several dynamical effects were described, including relaxation oscillations, hysteresis, etc.

To demonstrate the real-life applicability of the model, we considered a polariton gas in CdTe microcavity deriving a phase diagram for this model and localizing the regions of the condensed/normal phases of the system. Though not claiming full agreement with experiment, we expect the form of the phase diagram, in particular the existence of the bistable region, to be a general feature of condensates in quasiparticle systems with finite lifetime.

We see the main advantage of the proposed dynamical equation being its ability to naturally describe frequency-dependent gain and describe correct equilibrium behaviour (in sense of aligning condensate and reservoir chemical potentials) in finite-lifetime limit.

We expect the approach presented here with the structure of the memory kernel proposed to be useful for describing not only the mean field stationary states, but also dynamical, statistical properties far from equilibrating. This requires incorporating fluctuations in the model. Doing so and considering coherence build-up in the condensate is a subject of future work.

\section{Acknowledgements}
N.A.A. and Yu.E.L. acknowledge the support by the Russian Science Foundation grant No. {\href{https://rscf.ru/en/project/23-42-10010/}{23-42-10010}}. Part of the work devoted to CdTe microcavity polariton phase diagram evaluation was supported  by the  Foundation for the advancement of theoretical physics and mathematics “Basis”.

\onecolumn\newpage\appendix
\section{The self-energy term}
\subsection{Analytical calculations}
We start the discussion here using an expression from the main text (note that this expression is dimensionless, $\beta=1$ is considered):
\begin{align}
    \Im [\Sigma^+_{\bs q}(\omega)]=&{-}8g_0^2\left[e^{\omega-\mu}{-}1\right]\sum_{\bs q_1, \bs q_2}f_{\bs q_1}(1{+}f_{\bs q_1+\bs q_2-\bs q})f_{\bs q_2}\delta(\omega+\varepsilon_{\bs q_1+\bs q_2-\bs q}-\varepsilon_{\bs q_1}-\varepsilon_{\bs q_2}).
\end{align}
Considering quadratic spectrum $\varepsilon_{\bs q}=\frac{q^2}{2m}$ for reservoir particles and passing to continuum limit with dimensionless momentum $q\to q\lambda_{dB}$, we obtain:
\begin{align}\label{eq:sigmaretim_ready_to_integrate}
    \Im [\Sigma^+_{\bs q}(\omega)]{=}&{-}2\frac{\pi V_0^2}{(2\pi)^4\lambda^4_{dB}}\left[e^{\omega-\mu}{-}1\right]\int \!d^2\bs q_1\int \!d^2\bs q_2\delta\left(\omega{+}(\bs q_1{+}\bs q_2{-}\bs q)^2{-}\bs q_1^2{-}\bs q_2^2\right) f(q^2_1)\left[1{+}f\left(\left(\bs q_1{+}\bs q_2-\bs q\right)^2\right)\right]f\left(q_2^2\right)
\end{align}
It's now helpful to use $q_{+}=\frac{\bs q_1+\bs q_2}2$ and $q_{-}=\bs q_1-\bs q_2$ as integration variables:
\begin{align}
    \Im [\Sigma^+_{\bs q}(\omega)]{=}{-}2\frac{\pi V_0^2}{(2\pi)^4\lambda^4_{dB}}\left[e^{\omega-\mu}{-}1\right]\int d^2\bs q_+\int d^2\bs q_-\delta&\left(2q_+^2{-}\frac{q^2_-}2{-}4\bs q\bs q_+{+}\omega{+}q^2\right) f_{B}\left[\left(\bs q_+{-}\frac{\bs q_-}2\right)^2\right]\nonumber\\
    &\times\left(1{+}f_B\left[\left(2\bs q_+{-}\bs q\right)^2\right]\right)f_{B}\left[\left(\bs q_+{+}\frac{\bs q_-}2\right)^2\right]
\end{align}
After integrating out the delta-function:
\begin{align}
    \Im [\Sigma^+_{\bs q}(\omega)]{=}{-}\frac{V_0^2}{(2\pi)^3\lambda^4_{dB}}\left[e^{\omega-\mu}{-}1\right]\int_0^{\infty}d^2\bs q_+\int_0^{\infty}d^2\bs n_-\Theta\left(2q_+^2{-}4\bs q\bs q_+{+}\omega{+}q^2\right)\left(1{+}f_B\left[\left(2\bs q_+{-}\bs q\right)^2\right]\right) \times\nonumber\\
    f_{B}\left[2q_+^2{-}2\bs q\bs q_+{+}\frac{\omega}2{+}\frac{q^2}2{-}\bs n_-\bs q_+\sqrt{4q_+^2{-}8\bs q\bs q_+{+}2\omega{+}2q^2}\right]f_{B}\left[2q_+^2{-}2\bs q\bs q_+{+}\frac{\omega}2{+}\frac{q^2}2{+}\bs n_-\bs q_+\sqrt{4q_+^2{-}8\bs q\bs q_+{+}2\omega{+}2q^2}\right]
\end{align}
Here $\bs n_{-}=\frac{\bs q_-}{q_-}$ is a unitary 2D vector. We now use $\bs q_{\delta}=\bs q_+{-}\bs q$ and the corresponding unitary vector $\bs n_{\delta}$ to obtain:
\begin{align}
    &\Im [\Sigma^+_{\bs q}(\omega)]{=}{-}\frac{V_0^2}{(2\pi)^3\lambda^4_{dB}}\left[e^{\omega-\mu}{-}1\right]\int_0^{\infty}d^2\bs q_{\delta}\int_0^{\infty}d^2\bs n_-\Theta\left(2 q_{\delta}^2{-}q^2{+}\omega\right)\left(1{+}f_B\left[\left(2\bs q_{\delta}{+}\bs q\right)^2\right]\right) \times\nonumber\\
    &f_{B}\left[2\bs q_{\delta}(\bs q{+}\bs q_{\delta}){+}\frac{\omega}2{+}\frac{q^2}2{-}\bs n_-(\bs q{+}\bs q_{\delta})\sqrt{4 q_{\delta}^2{-}2q^2{+}2\omega}\right]f_{B}\left[2\bs q_{\delta}(\bs q{+}\bs q_{\delta}){+}\frac{\omega}2{+}\frac{q^2}2{+}\bs n_-(\bs q{+}\bs q_{\delta})\sqrt{4 q_{\delta}^2{-}2q^2{+}2\omega}\right]
\end{align}
The theta-function specifies the lower integration limit as follows:
\begin{align}
    \Im [\Sigma^+_{\bs q}(\omega)]{=}{-}\frac{V_0^2}{(2\pi)^3\lambda^4_{dB}}\left[e^{\omega-\mu}{-}1\right]\int d^2\bs n_{\delta}\int d^2\bs n_-\int_{\sqrt{\frac{q^2-\omega}{2}}}^{\infty}q_{\delta} d q_{\delta}\left(1{+}f_B\left[4q_{\delta}^2+q^2+4q_{\delta}\bs q\bs n_{\delta}\right]\right) \times\nonumber\\
    f_{B}\left[2q_{\delta}^2{+}2q_{\delta}\bs n_{\delta}\bs q{+}\frac{\omega{+}q^2}2{-}\bs n_-(\bs q{+}q_{\delta}\bs n_{\delta})\sqrt{4 q_{\delta}^2{-}2q^2{+}2\omega}\right]f_{B}\left[2q_{\delta}^2{+}2q_{\delta}\bs n_{\delta}\bs q{+}\frac{\omega{+}q^2}2{+}\bs n_-(\bs q{+}q_{\delta}\bs n_{\delta})\sqrt{4 q_{\delta}^2{-}2q^2{+}2\omega}\right]
\end{align}

Now introduce angles $\bs n_-\bs n_{q}=\cos(\alpha_-)$, $\bs n_{\delta}\bs n_{q}=\cos(\alpha_{\delta})$, $\bs n_-\bs n_{\delta}=\cos(\alpha_{\delta}-\alpha_-)$:
\begin{align}\label{eq:numint_source}
    &\Im [\Sigma^+_{\bs q}(\omega)]{=}{-}\frac{V_0^2}{(2\pi)^3\lambda^4_{dB}}\left[e^{\omega-\mu}{-}1\right]\frac12\int_0^{2\pi} d\alpha_{-}\int_0^{2\pi} d\alpha_{\delta}\int_{\sqrt{\frac{q^2-\omega}{2}}}^{\infty} dq_{\delta}q_{\delta}\left(1{+}f_B\left[4q_{\delta}^2+q^2+4q_{\delta}q\cos(\alpha_{\delta})\right]\right) \times\nonumber\\
    &f_{B}\left[2q_{\delta}^2{+}2q_{\delta}q\cos(\alpha_{\delta}){+}\frac{\omega{+}q^2}2{-}\left[q\cos(\alpha_{-}){+}q_{\delta}\cos(\alpha_{\delta}-\alpha_{-})\right]\sqrt{4 q_{\delta}^2{-}2q^2{+}2\omega}\right]\times\nonumber\\
    &f_{B}\left[2q_{\delta}^2{+}2q_{\delta}q\cos(\alpha_{\delta}){+}\frac{\omega{+}q^2}2{+}\left[q\cos(\alpha_{-}){+}q_{\delta}\cos(\alpha_{\delta}-\alpha_{-})\right]\sqrt{4 q_{\delta}^2{-}2q^2{+}2\omega}\right]=\frac{V_0^2}{(2\pi)^3\lambda^4_{dB}}I_{\bs q}(\omega)
\end{align}
This expression may be used for evaluating $I_{\bs q}(\omega)$. However, asymptotic behaviour may be studied analytically.

Considering $\omega\to\infty$:
\begin{align}
    I_{\bs q}(\omega{\to}\infty){=}{-}\left[e^{\omega-\mu}\right]\int_0^{2\pi} d\alpha_{-}\int_0^{2\pi} d\alpha_{+}\int_{0}^{\infty} dq_{+}q_{+}\frac{e^{\mu - \omega}}{e^{-\mu + 4q_+^2} - 1}={-}\frac{\pi^2}2\int_{0}^{\infty} dz\frac{1}{e^{-\mu + z} - 1}={-}\frac{\pi^2}2\ln(1-e^{\mu})
\end{align}

The opposite limit $\omega\to\infty$ (note that $q\sim 1$ is the thermal momentum magnitude):
\begin{align}
    I_{\bs q}(\omega{\to}{-}\infty){=}{-}\left[-1\right]\int_0^{2\pi} d\alpha_{-}\int_0^{2\pi} d\alpha_{\delta}\int_{\sqrt{-\frac{\omega}{2}}}^{\infty} dq_{\delta}q_{\delta}e^{2\mu - \omega - 4q^2_{\delta}}=4\pi^2\frac{e^{2\mu+\omega}}{8}=\frac{\pi^2}{2}e^{2\mu+\omega}.  
\end{align}

Note that these asymptotic expressions are independent of $q$. 

\subsection{Fitting the $I_0(\omega)$ curve}
We have numerically integrated the expression for $I_0(\omega)$ from $\eqref{eq:numint_source}$ and performed fitting of the left peak (as described in the main text) with a model expression. The results, as well as the approximate least-squares estimates which justify analytical approximations for the $\Lambda^+(\tilde n)$ and $\gamma(\tilde n)$ dependences are presented below in the Table \ref{table:fitting_data} and Fig. \ref{fig:OLS}:

\begin{table}[H]
 \pgfkeys{/pgf/number format/.cd, std, std=-3:2}
  \begin{center}
    \begin{filecontents*}{data/fitting.dat}
n	lnn	L	lnL	g	lng
0.02	-3.91202300542815	0.041	-3.19418321227783	0.386	-0.951917909517306
0.03	-3.50655789731998	0.0689	-2.67509910096252	0.3485	-1.05411704878156
0.05	-2.99573227355399	0.163	-1.81400507817537	0.344	-1.06711362160874
0.07	-2.65926003693278	0.249	-1.39030238251743	0.269	-1.3130438993803
0.1	-2.30258509299405	0.51	-0.673344553263766	0.226	-1.48722027970985
0.15	-1.89711998488588	0.992	-0.00803217169726427	0.15	-1.89711998488588
0.2	-1.6094379124341	1.349	0.299363577225619	0.09	-2.40794560865187
0.3	-1.20397280432594	1.864	0.6227247162634	0.029	-3.54045944899566
0.4	-0.916290731874155	2.05	0.717839793150317	0.0088	-4.73300355749798
0.5	-0.693147180559945	3.09	1.12817109090965	0.00715	-4.94064292227622
0.8	-0.22314355131421	4	1.38629436111989	0.000168	-8.69154657856103
\end{filecontents*}
\pgfplotstabletypeset[
    precision=2,        
    column type={|||},
    display columns/0/.style={precision=2}, 
    display columns/1/.style={precision=2},
    display columns/2/.style={precision=2},
    fixed,
    set thousands separator={},
    fixed zerofill,
	col sep=tab,
	column type=c,
	every head row/.style={before row=\hline,after row=\hline},
	every last row/.style={after row=\hline},	
        every first column/.style={column type/.add={|}{}},
	every column/.style={column type/.add={}{|}},
        columns/n/.style={column name={$\tilde n$}
        },
        columns/lnn/.style={column name={$\ln(\tilde n)$}
        },
        columns/L/.style={column name={$\Lambda^+$}
        },
        columns/lnL/.style={column name={$\ln(\Lambda^+)$}
        },
        columns/g/.style={column name={$\gamma$}
        },
        columns/lng/.style={column name={$\ln(\gamma)$}
        },
]{data/fitting.dat}
\end{center}
\caption{\label{table:fitting_data} The results of fitting the function $I_0(\omega)$.}
\end{table}

\begin{figure}[H]
\begin{minipage}{0.49\linewidth}
\begin{figure}[H]
\begin{center}
   \begin{tikzpicture}
    \begin{axis}[
        title=$\ln\Lambda^+$ as a function of $\ln(\tilde n)$,
        width=\linewidth,
        xlabel={$\ln(\tilde n)$},
        ylabel={$\ln(\Lambda^+)$},
        y tick label style={
        /pgf/number format/.cd,
            fixed,
            fixed zerofill,
            precision=0,
        /tikz/.cd
    },
    x tick label style={
        /pgf/number format/.cd,
            fixed,
            fixed zerofill,
            precision=1,
        /tikz/.cd
    },
        grid=both,
        grid style={line width=.1pt, draw=gray!30},
        major grid style={line width=.2pt,draw=gray!100},
        axis lines=middle,
        minor tick num=3,
        xmin = -4, xmax = 0, ymin = -4, ymax = 2,
        legend pos=south east
    ]
    \addplot + [only marks, black] table[x=lnn,y=lnL]{data/fitting.dat};
    \addlegendentry{Fitting results};
    \addplot [thick,blue] table[x=lnn, y={create col/linear regression={y=lnL}}]{data/fit1.dat};
    \addlegendentry{
        $\ln(\Lambda^+) =1.55 \cdot \ln(\tilde n)
        \pgfmathprintnumber[print sign]2.83$};
    \addplot [thick,red] table[x=lnn, y={create col/linear regression={y=lnL}}]{data/fit2.dat};
    \addlegendentry{
        $\ln(\Lambda^+) =0.83 \cdot \ln(\tilde n)\pgfmathprintnumber[print sign]1.60$};
    \end{axis}
    \end{tikzpicture} 
\end{center}
\end{figure}
\end{minipage}
\hfill
\begin{minipage}{0.49\linewidth}
\begin{figure}[H]
\begin{center}
   \begin{tikzpicture}
    \begin{axis}[
        title=$\ln\gamma$ as a function of $\tilde n$,
        width=\linewidth,
        xlabel={$\tilde n$},
        ylabel={$\ln(\gamma^+)$},
        y tick label style={
        /pgf/number format/.cd,
            fixed,
            fixed zerofill,
            precision=0,
        /tikz/.cd
    },
    x tick label style={
        /pgf/number format/.cd,
            fixed,
            fixed zerofill,
            precision=1,
        /tikz/.cd
    },
        grid=both,
        grid style={line width=.1pt, draw=gray!30},
        major grid style={line width=.2pt,draw=gray!100},
        axis lines=middle,
        minor tick num=3,
        xmin = 0, xmax = 0.9, ymin = -9, ymax = 0,
        legend pos= south west
    ]
    \addplot + [only marks, black] table[x=n,y=lng]{data/fitting.dat};
     \addplot [thick,blue] table[x=n, y={create col/linear regression={y=lng}}]{data/fitting.dat};
     \addlegendentry{
        $\ln(\gamma) =\pgfmathprintnumber{\pgfplotstableregressiona} \cdot \tilde n
        \pgfmathprintnumber[print sign]{\pgfplotstableregressionb}$};
        \label{graph1}
        \end{axis}
    \end{tikzpicture} 
\end{center}
\end{figure}
\end{minipage}
\caption{Fitting parameters ($\Lambda^+$ and $\gamma$) presented as functions of normalised density $\tilde n = n\lambda_{\rm db}^2$.}
\label{fig:OLS}
\end{figure}

\section{Stability analysis}
\subsection{Stationary points. Stability}
For stability analysis, we need to express equations in the form of an ODE system:
\begin{align}
    \dot V &= -\frac{V^2}{2}-\left(\frac{\Gamma}2+\gamma\right)V+2\nu(\nu - \varepsilon-ge^r)+\left(2\Lambda^+\frac{\mu}{\gamma}-\gamma\Gamma\right)\\
    \dot r &= V\\
    \dot \nu &= {-}\frac32V\nu{+}\left(\varepsilon{+}2g e^r\right)V{+}\left(\frac{V}{2}{-}\gamma\right)\left(\nu{-}\varepsilon{-}ge^r\right){-}\frac12\Gamma\nu{-}\Lambda^+.
\end{align}
We may linearise it by substitution $\nu = \nu_{\rm eq}+\delta \nu,\, V=\delta V,\,r=r_{\rm eq}+\delta r$.
Leaving first order terms only:
\begin{align}
    \dot V &= -\left(\frac{\Gamma}2+\gamma\right)V+2\delta\nu(\nu- \varepsilon-g\rho)+2\nu(\delta \nu -g\rho\delta r),\\
    \delta \dot r &= V,\\
    \delta\dot \nu &= {-}\frac32V\nu{+}\left(\varepsilon{+}2g \rho\right)V{-}\gamma\left(\delta \nu{-}g\rho\delta r\right){+}\left(\frac{V}{2}\right)\left(\nu{-}\varepsilon{-}g\rho\right){-}\frac12\Gamma\delta\nu.
\end{align}
Seeking the eigenmodes of this system in a form
$$
\begin{pmatrix}V\\\delta r\\ \delta \nu\end{pmatrix}=\begin{pmatrix}\delta V_0\\\delta r_0\\ \delta \nu_0\end{pmatrix}e^{\chi t},
$$
leads to the following characteristic equation:
\begin{align}
    \begin{vmatrix}
        -\left(\frac{\Gamma}2+\gamma\right)-\chi&-2g\rho\nu&2(2\nu- \varepsilon-g\rho)\\1&-\chi&0\\\left(\frac{\varepsilon}2{+}\frac32g \rho{-}\nu\right)&\gamma g\rho&{-}\left(\frac{\Gamma}2 {+}\gamma\right)-\chi
    \end{vmatrix}=0.
\end{align}
Denoting $f^2=\left(\gamma+\frac{\Gamma}2\right)^2{+}(\varepsilon-2\nu+g\rho)^2+2g\rho(\varepsilon-\nu+g\rho)$, we may rewrite it as a cubic equation with respect to the eigenvalue $\chi$:
\begin{align}
    \chi^3+ (2 \gamma +\Gamma )\chi ^2+f^2\chi  +2g\rho\left[\Gamma\nu+\Lambda^+\right]=0
\end{align}
Using expressions \eqref{eq:satationary_lower_upper} for the stationary points, one may simplify ($+$ sign in the last term corresponds to the upper stationary point):
\begin{align}
    \chi^3+ (2 \gamma +\Gamma )\chi ^2+f_{\pm}^2\chi  \pm2g\rho_{\pm}\sqrt{D}=0\label{eq:characteristic_complex}
\end{align}
with $f_{\pm}^2$ being expressed as:
\begin{align}
    f_{\pm}^2=\left(\gamma+\frac{\Gamma}2\right)^2+\frac{\left[(2 \gamma +\Gamma )\Lambda^+\mp(2 \gamma -\Gamma )\sqrt{D}\right]^2}{4 \gamma ^2 \Gamma ^2}+\frac{\text{g$\rho_{\pm}$}}{\gamma} \left({\Lambda^+}\pm{\sqrt{D}}\right).\label{eq:f2pm}
\end{align}
First note that whenever the lower stationary point exists ($D>0$ and $\rho_-\geq0$) the free term on the left-hand side of \eqref{eq:characteristic_complex} is negative. It is enough to conclude that the lower solution has at least one positive real eigenvalue $\chi>0$, leading to instability.

For the upper stationary point, all the terms are positive whenever it exists. Therefore, the cubic equation \eqref{eq:characteristic_complex} has one real negative eigenvalue and two complex-conjugate eigenvalues. We may explicitly consider the real and imaginary components of $\chi=\kappa+i\Omega$, which leads to the following equation system:
\begin{equation}
	\begin{cases}
		\kappa ^3+\left(\kappa ^2-\Omega^2\right) (2 \gamma +\Gamma )+\kappa  \left(f_{\pm}^2{-}3 \Omega ^2\right)\pm2g\rho_{\pm}\sqrt{D}=0\\
		\Omega\left[3\kappa^2-\Omega^2+2\kappa\left(2\gamma+\Gamma\right)+f_{\pm}^2\right]=0
\end{cases}
\end{equation}

The non-zero frequency satisfying the second equation is $\Omega^2=3\kappa^2+2\kappa\left(2\gamma+\Gamma\right)+f_{\pm}^2$. By substitution, we get:
\begin{align}
    \kappa ^3+ (2 \gamma +\Gamma ) \kappa ^2+\left(\left(\gamma+\frac{\Gamma}2\right)^2+\frac{f_{\pm}^2}4\right)\kappa +\frac14\left(\gamma+\frac{\Gamma}2\right)f_{\pm}^2\mp\frac{g\rho_{\pm}}4\sqrt{D}=0\label{eq:stab_real_part}
\end{align}

The free term here may be expressed as follows:
\begin{align*}
    \frac14\left(\gamma+\frac{\Gamma}2\right)f_{\pm}^2\mp\frac{g\rho}4\sqrt{D}=\frac14\left(\gamma{+}\frac{\Gamma}2\right)\left[\left(\gamma{+}\frac{\Gamma}2\right)^2{+}\frac{\left[(2 \gamma {+}\Gamma )\Lambda^+\mp(2 \gamma {-}\Gamma )\sqrt{D}\right]^2}{4 \gamma ^2 \Gamma ^2}{+}\frac{\text{g$\rho_{\pm}$}}{\gamma} \left({\Lambda^+}\pm{\sqrt{D}}\right)\right]\mp\frac{g\rho_{\pm}}4\sqrt{D}=\nonumber\\
    =\frac1{4}{\left(\gamma+\frac{\Gamma}2\right)^3}{+}\frac14\left(\gamma+\frac{\Gamma}2\right)\frac{\left[(2 \gamma {+}\Gamma )\Lambda^+\mp(2 \gamma {-}\Gamma )\sqrt{D}\right]^2}{4 \gamma ^2 \Gamma ^2}{+}\frac{g\rho_{\pm}}{8\gamma}\left[(2\gamma+\Gamma)\Lambda^+\pm\Gamma\sqrt{D}\right].
\end{align*}
We may now note that all the coefficients of the cubic equation \eqref{eq:stab_real_part} are positive for $+$ sign chosen, which leads to $\kappa<0$ and allows to finally conclude that the upper stationary point is stable.

\subsection{Decaying solutions}
We start from the characteristic equations for the decaying solution, presented in the main text:
\begin{equation}	
	\begin{cases}
		\frac{\kappa^2}{2}=-\left(\frac{\Gamma}{2}+\gamma\right)\kappa+\left(2\frac{\mu}{\gamma}\Lambda^+-\gamma\Gamma\right)+2\Omega(\Omega-\varepsilon),\\
		0=-\frac{3}{2}\kappa\Omega+\varepsilon\kappa+\left(\frac{\kappa}{2}-\gamma\right)(\Omega-\varepsilon)-\frac{1}{2}\Gamma\Omega-\Lambda^+.
	\end{cases}
	\label{eq:decay_stability_appendix}
\end{equation}
	After excluding $\nu$ from these equations and using a substitution $\kappa \to -\gamma -\frac{\Gamma }{2}+y$:
	\begin{align}
        \Omega&=\frac{\varepsilon\left(\gamma{+}\frac{\kappa}{2}\right){-}\Lambda^+}{\kappa{+}\gamma{+}\frac{\Gamma}2},\\
	    f(y)&=y^4{+}\left(\epsilon ^2{-}\left(\gamma{-}\frac{\Gamma}{2}\right)^2{-}4 \Lambda^+ \frac{\mu}{\gamma} \right)  y^2 {-}  \left(\left(\gamma{-}\frac{\Gamma}{2}\right) \epsilon {-} 2\Lambda^+ \right)^2{=}0.
	\end{align}
	Since the last term is given by subtraction of a non-negative perfect square, in order for all the roots to be less than $\gamma+\frac{\Gamma}2$ (which is the same as requesting all the $\kappa$s to be negative), we need to impose a condition $f\left(\frac{\Gamma}{2}+\gamma\right)\geq0$. By direct substitution, one may verify that this stability condition is equivalent to the inequality
	\begin{align}
	    \left(\frac{\rho_++\rho_-}2\right)^2\geq\left(\frac{\rho_+-\rho_-}2\right)^2=\left(\frac1{\Gamma}+\frac1{2\gamma}\right)^2\frac{D}{g^2} 
	\end{align}
	or $\rho_+\rho_-\geq0$. From the expression above, we see that $D<0$ (when there are no stationary points at all) also satisfies the stability condition.
	The eigenvalues themselves are of the following form:
	\begin{align}
	    \kappa_{\pm}={-}\gamma{-}\frac{\Gamma}2{\pm}\frac{\sqrt 2\sqrt{\sqrt{\left(\epsilon ^2{-}\left(\gamma{-}\frac{\Gamma}{2}\right)^2{-}4 \Lambda^+  \frac{\mu}{\gamma}\right)^2+4\left(\left(\gamma{-}\frac{\Gamma}{2}\right) \epsilon {-} 2\Lambda^+ \right)^2}{-}\left(\epsilon ^2{-}\left(\gamma{-}\frac{\Gamma}{2}\right)^2{-}4 \Lambda^+  \frac{\mu}{\gamma} \right)}}2=-\gamma-\frac{\Gamma}2\pm\frac{\Delta\kappa}{2}.
	\end{align}
The corresponding frequencies are:
\begin{align}
    \Omega_{\pm}{=}\frac{\varepsilon\left(\gamma{+}\frac{\kappa}{2}\right){-}\Lambda^+}{\kappa{+}\gamma{+}\frac{\Gamma}2}=\frac{\varepsilon}{2}\mp\frac{\varepsilon\left(\gamma{-}\Gamma\right)-2\Lambda^+}{\Delta\kappa}.
\end{align}

\end{document}